\documentclass[pre,twocolumn,amsmath,amssymb,showpacs]{revtex4}
\usepackage{graphicx}% Include figure files
\usepackage{dcolumn}% Align table columns on decimal point
\usepackage{bm}% bold math
\begin{document}
\baselineskip 0.423cm
%---------------------------------------------------------------------
%
\newcommand{\vspfigA}{\vspace{0cm}}  
\newcommand{\vspfigB}{\vspace{0cm}} 
\newcommand{\vspfigC}{\vspace{0.3cm}}
\newcommand{\widthfig}{0.4\textwidth}
\newcommand{\transP}[5]{
   {#1}\!\left(\begin{array}{c} \!\! #2 \!\! \\ \!\! #3 \!\! 
  \end{array}\right| \left. \begin{array}{c}
   \!\! #4 \!\! \\ \!\! #5 \!\! \end{array}\right) }
\newcommand{\transS}[5]{
   {#1}\!(\begin{array}{c} \!\! \scriptstyle{#2} \!\! \\ 
   \!\! \scriptstyle{#3} \!\! \end{array} |
  \begin{array}{c} \!\! \scriptstyle{#4} \!\! \\ 
  \!\! \scriptstyle{#5} \!\! \end{array}) }
\newcommand{\pathaveA}[1]{
   \left\langle\!\!\!\left\langle #1
   \right\rangle\!\!\!\right\rangle_{t} }
\newcommand{\spaEq}{\hspace{0.8cm}}
\newcommand{\ym}{\tilde{y}^{*}\!\!}
\newcommand{\ymd}{\dot{\ym}{\,}}

%%%%%%%%%%%%%%%%%%%%%%%%%%%%%%%%%%%%%%%%%%%%%%%%%%%%%%%%%%%%%%%%%%%%%%

\title{Onsager-Machlup theory for nonequilibrium steady 
   states and fluctuation theorems}
\author{Tooru Taniguchi and E. G. D. Cohen}
\affiliation{The Rockefeller University, 1230 York Avenue, 
New York, NY 10021, USA.} 
\date{\today}

\begin{abstract}
   A generalization of the Onsager-Machlup theory from 
equilibrium to nonequilibrium steady states 
and its connection with recent fluctuation theorems are discussed  
for a dragged particle restricted by a harmonic potential 
in a heat reservoir. 
   Using a functional integral approach, the probability functional 
for a path is expressed in terms of a Lagrangian function 
from which an entropy production rate and dissipation functions 
are introduced, and nonequilibrium thermodynamic relations 
like the energy conservation law and the second law of thermodynamics 
are derived. 
   Using this Lagrangian function we establish two nonequilibrium 
detailed balance relations, which not only lead to 
a fluctuation theorem for work but also to one related 
to energy loss by friction. 
   In addition, we carried out the functional integrals for heat 
explicitly, leading to the extended fluctuation theorem for heat. 
  We also present a simple argument for this extended  
fluctuation theorem in the long time limit. 
\end{abstract}
\maketitle

%%%%%%%%%%%%%%%%%%%%%%%%%%%%%%%%%%%%%%%%%%%%%%%%%%%%%%%%%%%%%%%%%%%%%%
\section{Introduction}

   Fluctuations play an important role in descriptions 
of nonequilibrium phenomena. 
   A typical example is the fluctuation-dissipation theorem, 
which connects transport coefficients to fluctuations 
in terms of auto-correlation functions. 
   This theorem can be traced back to Einstein's relation \cite{E05}, 
Nyquist's theorem \cite{J28,N28}, Onsager's arguments for  
reciprocal relations \cite{O31a,O31b,C45}, etc., 
and it was established in linear response theory 
in nonequilibrium statistical mechanics near equilibrium 
\cite{G51,CW51,K57}.  
   Another example of fluctuation theories 
is Onsager-Machlup's fluctuation theory around equilibrium 
\cite{H52,OM53,MO53}.
   It is characterized by the usage of a functional integral 
technique  
for stochastic linear relaxation processes, 
and leads to a variational principle  
known as Onsager's principle of minimum energy dissipation.  
   Many efforts have been devoted to 
obtain a generalization, 
for example, to the cases of nonlinear dynamics \cite{H76,H77,Y71,HR81,R89} 
and nonequilibrium steady states \cite{BSG01,BSG02,G02}. 

   Recently, another approach to fluctuation theory leading to  
fluctuation theorems has drawn considerable attention 
in nonequilibrium statistical physics 
\cite{ECM93,ES94,GC95}.   
   They are asymmetric relations for the distribution functions for 
work, heat, etc., and they may be satisfied 
even in far from equilibrium states or for non-macroscopic systems 
which are beyond conventional statistical thermodynamics. 
%   These theorems are purely for nonequilibrium processes. 
%so that it cannot be derived as  
%equilibrium statistical mechanics. 
   Originally they were proposed for deterministic chaotic dynamics, 
but they can also be justified for stochastic systems 
\cite{K98,LS99,C99,C00}. 
   Moreover, laboratory experiments to check these 
fluctuation theorems 
have been made \cite{CL98,WSM02,CGH04,FM04,GC05,SST05}. 
  
   From our accumulated knowledge on fluctuations, 
it is meaningful to ask for relations among the different 
fluctuation theories. 
   It is already known that the fluctuation-dissipation theorem, 
as well as Onsager's reciprocal relations, can be derived from 
fluctuation theorems near equilibrium states \cite{ECM93,LS99,G96}. 
   The heat fluctuation theorem can also be regarded 
as a refinement of the second law of thermodynamics. 
%   However, the relation 
%between Onsager-Machlup's fluctuation theory and 
%the steady state fluctuation theorems has not been explored so far. 
%   Besides, it has not been very obvious what is the physical origin 
%to justify fluctuation theorems, with leaving unclear how far 
%from equilibrium states or how small systems 
%fluctuation theorems can be satisfied.  

   The principal aims of this paper are twofold. 
   First, we generalize Onsager and Machlup's original 
fluctuation theory around equilibrium 
to fluctuations around nonequilibrium steady states 
using the functional integral approach. 
   For this nonequilibrium steady state Onsager-Machlup theory 
we discuss the energy conservation law (i.e. the analogue of the 
first law of thermodynamics), the second law of thermodynamics, 
and Onsager's principle of minimum energy dissipation. 
   As the second aim of this paper, we discuss fluctuation 
theorems based on our generalized Onsager-Machlup theory. 
   Since the systems we consider are in a nonequilibrium steady 
state, the equilibrium detailed balance condition is violated. 
   We propose generalized forms of the detailed balance conditions  
for nonequilibrium steady states, which we call  
\emph{nonequilibrium detailed balance relations}, and 
show that the fluctuation theorem for work 
can be derived from it. 
   To demonstrate the efficacy of 
nonequilibrium detailed balance as 
an origin of fluctuation theorems, we also show another 
form of nonequilibrium detailed balance, which leads to 
another fluctuation theorem for energy loss by friction. 
   We also show how a heat fluctuation theorem can be 
derived from our generalized Onsager-Machlup theory, 
by carrying out explicitly a functional integral 
and reducing its derivation to a previous one 
discussed in Refs. \cite{ZC03a,ZC04}.  
   In addition, we give a simple argument 
leading to the long-time ($t\rightarrow+\infty$) fluctuation 
theorem for heat, based on the independence between the work 
distribution and the energy-difference distribution.    
   
   In this paper, in order to make our arguments 
as concrete and simple as possible, we apply 
our theory to a specific nonequilibrium Brownian particle model 
described by a Langevin equation (cf. \cite{S98}). 
   It has been used to discuss fluctuation theorems 
\cite{MJ99,TTM02,ZC03a,ZC03b,ZC04}, and also to 
describe laboratory experiments for a Brownian particle captured 
in an optical trap which moves with a constant velocity 
through a fluid \cite{WSM02,TTM02}, 
as well as  for an electric circuit 
consisting of a resistor and capacitor \cite{ZCC04,GC05}. 

   The outline of this paper is as follows. 
   In Sec. \ref{DraggedBrownianParticle}, we introduce our model 
and give some of its properties using a functional integral 
approach. 
   In Sec. \ref{OnsagerMachlupSteadyStates}, 
we discuss a generalization of Onsager-Machlup's fluctuation theory
to nonequilibrium steady states, and obtain 
the energy conservation law,  
the second law of thermodynamics, 
i.e. a nonequilibrium steady state thermodynamics, 
and Onsager's principle of minimum energy dissipation 
for such states. 
   In Sec. \ref{FluctuationTheoremWork}, 
we introduce the concept of nonequilibrium detailed balance, 
and obtain a fluctuation theorem for work from it. 
   In Sec. \ref{FluctuationTheoremFriction}, 
we discuss another type of nonequilibrium detailed balance, 
which leads to a fluctuation theorem for energy loss by friction.
   In Sec. \ref{FluctuationTheoremHeat}, 
we sketch a derivation of a fluctuation theorem for heat 
by carrying out a functional integral 
and reducing it to the previous derivation 
\cite{ZC03a,ZC04}. 
  In addition, we give a simple argument 
for the heat fluctuation theorem 
in the long time limit. 
   In Sec. \ref{InertiaEffects}, we briefly discuss  
inertial effects on the fluctuation theorems, which lead to 
four new fluctuation theorems. 
   In Sec. \ref{ConclusionsRemarks}, we summarize our results 
in this paper and discuss some consequences of them.

%%%%%%%%%%%%%%%%%%%%%%%%%%%%%%%%%%%%%%%%%%%%%%%%%%%%%%%%%%%%%%%%%%%%%%
\section{Dragged Particle in a Heat Reservoir}
\label{DraggedBrownianParticle}

\begin{figure}[!t]
\vspfigA
\begin{center}
\includegraphics[width=\widthfig]{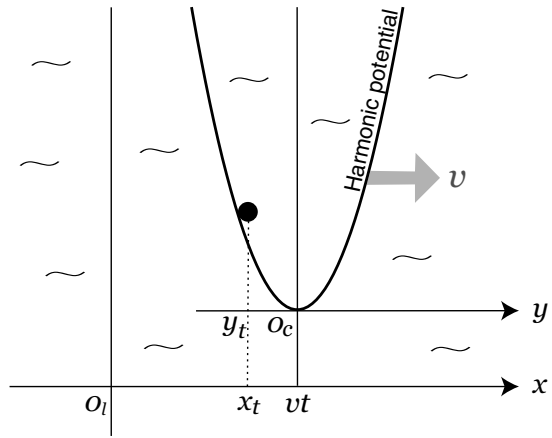}
\vspfigB
\caption{Schematic illustration for a particle 
   trapped by a harmonic potential 
   dragged with a constant velocity $v$ in a reservoir. 
%      The work $W$ is required to drag the particle 
%   and to keep the system 
%   in a nonequilibrium steady state, and the heat $Q$ is 
%   released by a coupling of the particle 
%   to the heat reservoir. 
%      The stable position of the particle is shifted 
%   from the bottom of the harmonic potential 
%   as an effect to drag the particle. 
      Here, $x$ and $y$ ($0_{l}$ and $0_{c}$) 
   are the axes (the origins) for the laboratory 
   and comoving frame, respectively, in the direction 
   of the motion of the particle. 
      The particle is at the position $y_{t}$ 
   ($x_{t}$) at time $t$ in the comoving (laboratory) frame, 
   respectively, which are related by $y_{t}=x_{t}-vt$. 
      After the relaxation time $\tau$, the system will reach a 
   nonequilibrium steady state.}
\label{figA1system}
\end{center}
\vspfigC
\end{figure}

   The system considered in this paper is a particle 
dragged by a constant velocity $v$ in a fluid as a heat reservoir.   
   The dynamics of this system is expressed as a 
Langevin equation 
\cite{noteIIa}
\begin{eqnarray}
   m\frac{d^{2} x_{t}}{dt^{2}} 
   = - \alpha \frac{d x_{t}}{dt} 
     - \kappa \left(x_{t}-v t\right) + \zeta_{t} 
\label{LangeEq3}\end{eqnarray}
for the particle position $x_{t}$ at time $t$ in the laboratory 
frame. 
   Here, $m$ is the particle mass, and 
on the right-hand side of Eq. (\ref{LangeEq3}) 
the first term is the friction force with the friction 
constant $\alpha$, the second term is the harmonic 
potential force with the spring constant $\kappa$ 
to confine the particle, and the third term, 
due to the coupling to the heat reservoir,   
is a Gaussian-white noise $\zeta_{t}$, 
whose first two auto-correlations are given by   
\begin{eqnarray}
   \langle \zeta_{t} \rangle &=& 0 ,
      \label{RandomForce1} \\
   \langle \zeta_{t_{1}}\zeta_{t_{2}} \rangle 
      &=& \frac{2\alpha}{\beta} \delta(t_{1}-t_{2})
      \label{RandomForce2}
\end{eqnarray}
with the inverse temperature $\beta$ of the reservoir 
and the notation $\langle \cdots\rangle$ 
for an initial ensemble average.  
%   The term $m d^{2} x_{t}/dt^{2}$ on the left-hand side 
%of Eq. (\ref{LangeEq3}) is the inertial term. 
%Eq. (\ref{LangeEq3}) is attributed into 
%the Langevin equation (\ref{LangeEq1}) under the over-damped 
%assumption by neglecting this inertial term. 
   The coefficient $2\alpha/\beta$ in Eq. (\ref{RandomForce2}) 
is determined by the fluctuation-dissipation theorem,   
so that in the case $v=0$  
the stationary state distribution function 
for the dynamics (\ref{LangeEq3}) 
is expressed by a canonical distribution.  
   A schematic illustration for this system is given in Fig. 
\ref{figA1system}.
%
%---------------------------------------------------------------------
%--------------------------------------------------------------------- 

   In this paper, except in Sec. \ref{InertiaEffects}, 
we consider the over-damped case in which we neglect 
the inertial term $m d^{2}x_{t}/dt^{2}$, or assume simply  
an negligible small mass $m$. 
%   This assumption means 
% the neglect of fast movements of the particle, or simply, 
%an extremely small mass $m$. 
   Under this over-damped assumption, 
the Langevin equation (\ref{LangeEq3}) can be written as 
\begin{eqnarray}
   \frac{d x_{t}}{dt} = - \frac{1}{\tau} \left(x_{t}-v t\right) 
   + \frac{1}{\alpha} \zeta_{t}
\label{LangeEq1}\end{eqnarray}
with the relaxation time $\tau$ given by $\tau\equiv\alpha/\kappa$. 
%   Equation (\ref{LangeEq1}) comes from the Langevin equation 
%for the friction force $-\alpha dx_{t}/dt$ and 
%the harmonic force $-\kappa (x_{t}-v t)$ and the random force 
%$\zeta_{t}$, neglecting the inertial term $m d^{2}x_{t}/dt^{2}$ 
%with the particle mass $m$, namely, under the over-damping assumption. 
% 

   Equation (\ref{LangeEq1}) is for the position $x_{t}$ 
in the laboratory frame. 
   On the other hand, it is often convenient 
or simpler to discuss the nonequilibrium dynamics 
in the comoving frame \cite{TM04,ZC03b}. 
   The position $y_{t}$ in the comoving frame 
for the particle in our model is simply introduced as 
\begin{eqnarray}
   y_{t} \equiv x_{t} - v t. 
\label{ComovPosit1}\end{eqnarray}
   Using this position $y_{t}$, Eq. (\ref{LangeEq1}) can be  
rewritten as 
\begin{eqnarray}
   \frac{d y_{t}}{dt} = -\frac{1}{\tau} y_{t} - v 
   + \frac{1}{\alpha} \zeta_{t} ,
\label{LangeEq2}\end{eqnarray}
whose dynamics is invariant under the change $y_{t}\rightarrow -y_{t}$ 
and $v\rightarrow -v$, noting that the Gaussian-white noise  
property of $\zeta_{t}$ is not changed into  
$\zeta_{t}\rightarrow-\zeta_{t}$. 
   Note that in the comoving Langevin equation (\ref{LangeEq2}) 
there is no explicit $t$-dependent term in the dynamical equation, 
while the laboratory Langevin equation (\ref{LangeEq1}) 
has a $t$-dependence through the term $vt$, 
meaning Eq. (\ref{LangeEq2}) to be a little simpler than 
Eq. (\ref{LangeEq1}). 
   The constant term $-v$ in Eq. (\ref{LangeEq2}) expresses 
all effects of the nonequilibrium steady state in this model. 

   The system described by the Langevin equation 
(\ref{LangeEq2}), or equivalently Eq. (\ref{LangeEq1}), 
approaches a nonequilibrium steady state, because 
the particle will, for $t>\tau$, move steadily 
due to the external force that drags it through the fluid.  
   This force is 
given by $-\kappa y_{t}$, so the work rate 
$\dot{\mathcal{W}}^{(v)}(y_{t})$ to keep 
the particle in a steady state is expressed as 
\begin{eqnarray}
   \dot{\mathcal{W}}^{(v)}(y_{t}) = - \kappa y_{t} v. 
\label{WorkRate1}\end{eqnarray}
   We note that since $\dot{\mathcal{W}}^{(0)}(y_{t})=0$ for $v=0$, 
i.e. for the equilibrium state considered by 
Onsager and Machlup,  there is no work done, 
while in the nonequilibrium steady state for $v\neq 0$ 
work is done \cite{note2a}.
%   Noting $\dot{\mathcal{W}}^{(0)}(y_{t})=0$, 
%this characteristic distinguishes 
%the non-equilibrium steady state case $v\neq 0$ and 
%the relaxation case $v=0$ to an equilibrium state physically 
%\cite{note2a}.  
 
   We consider the transition probability 
$\transS{F}{y_{t}}{t}{y_{0}}{t_{0}}$ of the particle 
from $y_{0}(\equiv y_{t_{0}})$ at time $t_{0}$ 
to $y_{t}$ at time $t$, which is introduced as a 
transition integral kernel for the probability distribution 
$f(y_{t},t)$ at the position $y_{t}$ at time $t$ as 
\begin{eqnarray}
   f(y_{t},t) = \int dy_{0} \; \transP{F}{y_{t}}{t}{y_{0}}{t_{0}}
   f(y_{0},t_{0})
\label{ProbaDistrY1}\end{eqnarray}
with the initial distribution $f(y_{0},t_{0})$.  
%   Here, the probability distribution $f(y_{t},t)$ is  
%the solution of the corresponding Fokker-Planck equation 
%\cite{R89}.
   We can use various analytical techniques, 
for example the Fokker-Planck equation,   
whose solution gives the probability distribution $f(y_{t},t)$ 
\cite{K92,R89}, 
to analyze the transition probability for the dynamics 
expressed by the Langevin equation (\ref{LangeEq2}).
   As one such technique, 
motivated by Ref. \cite{OM53,MO53}, 
we use in this paper the functional integral technique \cite{R89}.  
   Using this technique, the transition probability 
is represented as 
\begin{eqnarray}
   \transP{F}{y_{t}}{t}{y_{0}}{t_{0}} 
   = \int_{y_{0}}^{y_{t}}
   \mathcal{D}y_{s} \; \exp\left[ \int_{t_{0}}^{t}ds\; 
   L^{(v)}\!\left(\dot{y}_{s},y_{s}\right)\right]
\label{TransProba1}\end{eqnarray}
where $L^{(v)}\!\left(\dot{y}_{s},y_{s}\right)$ is 
the Lagrangian function for this stochastic process, defined by
\begin{eqnarray}
    L^{(v)}\!\left(\dot{y}_{s},y_{s}\right) \equiv 
   -\frac{1}{4 D} 
   \left( \dot{y}_{s} +\frac{1}{\tau} y_{s} +v \right)^{2} ,
\label{Lagra1}\end{eqnarray} 
where 
%$\dot{y}_{t}$ is the time-derivative of $y_{t}$, and 
$D$ is the diffusion constant given by 
the Einstein relation $D \equiv 1/(\alpha\beta)$.
   [We outline a derivation of Eq. (\ref{TransProba1}) 
from Eq. (\ref{LangeEq2}) in Appendix 
\ref{TransitionProbabilityFunctionalIntegral}.]
   Here, the functional integral on the right-hand side of 
Eq. (\ref{TransProba1}) is introduced as  
\begin{eqnarray}
   &&\hspace{-0.8cm}
   \int_{y_{0}}^{y_{t}} 
   \mathcal{D}y_{s} \;X_{t}(\{y_{s}\}) 
   \nonumber \\
   &&
   = \lim_{N\rightarrow+\infty} 
      \left(\frac{1}{4\pi D\Delta t_{N}}\right)^{N/2}
   \nonumber \\
   && \spaEq \times 
      \int dy_{t_{N-1}} \int dy_{t_{N-2}} \cdots \int dy_{t_{1}} \;
      X_{t}(\{y_{s}\}) 
   \nonumber \\
\label{FunctInteg1}\end{eqnarray}
for any functional $X_{t}(\{y_{s}\})$,   
with $t_{n}\equiv t_{0} + n \Delta t_{N}$, $n=1,2,\cdots,N$,  
$\Delta t_{N}\equiv (t-t_{0})/N$,  
the initial time $t_{0}$, the final time $t_{N}=t$, 
the initial position $y_{0}$, and 
the final position $y_{t}$. 
   Here, we use the symbol $\{y_{s}\}$ 
in $X_{t}(\{y_{s}\})$ 
to show that $X_{t}(\{y_{s}\})$ is a functional 
of $\{y_{s}\}$ with $s\in[t_{0},t]$.
   It is important to note that from the representation 
(\ref{TransProba1}) of the transition probability 
$\transS{F}{y_{t}}{t}{y_{0}}{t_{0}}$ the functional 
$\exp[ \int_{t_{0}}^{t}ds\;L^{(v)}(\dot{y}_{s},y_{s})]$ 
can be regarded as the probability functional density  
of the path $\{y_{s}\}_{s\in[t_{0},t]}$. 

   For the Lagrangian function (\ref{Lagra1}), 
the functional integral 
on the right-hand side of Eq. (\ref{TransProba1}) 
can actually be carried out using Eq. (\ref{FunctInteg1}), 
to obtain, by a simple generalization of the 
well-known equilibrium ($v=0$) case,  
%
%\begin{widetext}
\begin{eqnarray}
   \transP{F}{y_{t}}{t}{y_{0}}{t_{0}} 
   & =&
%   =
      \frac{1}{\sqrt{4\pi D\mathcal{T}_{t}}}
      \nonumber \\
   &&\times  
      \exp\left\{-
      \frac{\left[y_{t}+v\tau -\left(y_{0}+v\tau\right)
      b_{t}\right]^{2}}
      {4 D\mathcal{T}_{t}}\right\} ,
      \nonumber \\
\label{TransProba2}\end{eqnarray}
%\end{widetext}
%
where $b_{t}$ and $\mathcal{T}_{t}$ are defined by 
$b_{t}\equiv \exp[-(t-t_{0})/\tau]$ 
and  $\mathcal{T}_{t} \equiv (\tau/2) (1 - b_{t}^{2}) $ 
so that $\mathcal{T}_{t} = t-t_{0} +\mathcal{O}((t-t_{0})^{2})$ 
\cite{noteIIA}. 
   Equation (\ref{TransProba2}) is simply a well known form 
of the transition 
probability for the Smoluchowski process \cite{R89}. 
   Inserting Eq. (\ref{TransProba2}) into  
Eq. (\ref{ProbaDistrY1}), 
using the normalization condition 
$\int dy_{0}\;f(y_{0},t_{0})=1$, 
and taking 
the limit $t\rightarrow+\infty$,    
we can show that 
for an arbitrary initial distribution $f(y_{0},t_{0})$, 
the probability distribution $f(y_{t},t)$ approaches 
to a nonequilibrium steady state (ss) distribution: 
\begin{eqnarray}
   f_{ss}(y_{t}) \equiv \lim_{t\rightarrow+\infty}f(y_{t},t) 
   = f_{eq}\left(y_{t}+v\tau\right) 
\label{SteadSolut1}\end{eqnarray}
in the long time limit. 
   Here, $f_{eq}(y)$ is the equilibrium distribution function 
given by 
\begin{eqnarray}
   f_{eq}(y) = \sqrt{\frac{\kappa\beta}{2\pi}} 
      \exp\left[-\beta U(y)\right]
\label{EquilDistr1}\end{eqnarray}
with the harmonic potential energy 
$U(y)\equiv \kappa y^{2}/2$.
   Equation (\ref{SteadSolut1}) implies that the steady state 
distribution $f_{ss}(y)$ is simply given by 
the equilibrium canonical distribution $f_{eq}(y)$ by shifting 
the position $y$ to $y+v\tau$.  
% as illustrated in Fig. \ref{figA1system}.
   [Note that there is no kinetic energy term in the canonical 
distribution (\ref{EquilDistr1}) under the over-damped assumption.]
   Equation (\ref{SteadSolut1}) implies that 
the average position of the particle is shifted 
from the bottom $y=0$ of the harmonic potential 
in the equilibrium state
to the position $y=-v\tau$ 
in the nonequilibrium steady state. 

   The functional integral approach has already been used 
to describe relaxation processes to thermal equilibrium  
with fluctuations and averages 
by Onsager and Machlup 
\cite{OM53,MO53}.
   In the next section, we generalize their argument 
to non-equilibrium steady states for our model, and 
construct a nonequilibrium steady state thermodynamics. 
   The results in Refs. \cite{OM53,MO53} can always be 
reproduced from our results 
in Sec. \ref{OnsagerMachlupSteadyStates} by taking $v=0$, 
i.e. in the equilibrium case. 
   In this generalization, we 
   %integrate 
determine the work to 
sustain the nonequilibrium steady state in the 
Onsager-Machlup theory, and also give a direct 
connection between the entropy production rate in 
the Onsager-Machlup theory and the heat discussed in Ref. 
\cite{ZC03a,ZC04}.

%%%%%%%%%%%%%%%%%%%%%%%%%%%%%%%%%%%%%%%%%%%%%%%%%%%%%%%%%%%%%%%%%%%%%%
\section{Onsager-Machlup Theory for Nonequilibrium Steady states}
\label{OnsagerMachlupSteadyStates}

   In the generalized Onsager-Machlup theory,  
%\cite{OM53,MO53,TM57}, 
the Lagrangian 
$L^{(v)}\!\left(\dot{y}_{s},y_{s}\right)$ 
can be written in the form
\begin{eqnarray}
   L^{(v)}(\dot{y}_{s},y_{s}) = -\frac{1}{2k_{B}}\left[ 
   \Phi^{(v)} (\dot{y}_{s}) + \Psi(y_{s}) 
   - \dot{\mathcal{S}}^{(v)}(\dot{y}_{s},y_{s}) \right]
   \nonumber \\
\label{Lagra2}\end{eqnarray}
where $k_{B}$ is the Boltzmann constant, and 
$\Phi^{(v)} (\dot{y}_{s})$, $\Psi(y_{s})$ 
and $\dot{\mathcal{S}}^{(v)}(\dot{y}_{s},y_{s})$ are defined by 
\begin{eqnarray}
   \Phi^{(v)} (\dot{y}_{s}) &\equiv& 
      \frac{\alpha}{2T}(\dot{y}_{s}+v)^{2} ,
      \label{DissiFunctA1} \\
   \Psi(y_{s}) &\equiv& 
      \frac{\alpha}{2T} \left(\frac{y_{s}}{\tau}\right)^{2} ,
      \label{DissiFunctB1} \\
   \dot{\mathcal{S}}^{(v)}(\dot{y}_{s},y_{s}) &\equiv& 
      -\frac{1}{T}\kappa y_{s}(\dot{y}_{s}+v) ,
      \label{EntroProdu1}
\end{eqnarray}
respectively, with the temperature $T\equiv (k_{B}\beta)^{-1}$. 
   These functions $\Phi^{(v)} (\dot{y}_{s})$ and $\Psi(y_{s})$ 
are called dissipation functions, 
%the corresponding function of force, 
while we call $\dot{\mathcal{S}}^{(v)}(\dot{y}_{s},y_{s})$ 
the entropy production rate. 
   In the next subsections \ref{HeatEntropyBalance} 
and \ref{DissipationFunctions}, we discuss the physical meaning 
of these quantities, and justify their names.

%---------------------------------------------------------------------
\subsection{Heat and energy balance equations}
\label{HeatEntropyBalance}
 
   Using the entropy production rate 
$\dot{\mathcal{S}}^{(v)}(\dot{y}_{s},y_{s})$, 
we introduce the heat $\mathcal{Q}^{(v)}_{t}(\{y_{s}\})$ 
produced by the system in 
the time-interval $[t_{0},t]$ as 
\begin{eqnarray}
   \mathcal{Q}^{(v)}_{t}(\{y_{s}\}) \equiv T \int_{t_{0}}^{t}ds\; 
   \dot{\mathcal{S}}^{(v)}(\dot{y}_{s},y_{s}) . 
\label{Heat1}\end{eqnarray}
%  
%(Here, we use the symbol $\{y_{s}\}$ 
%in $\mathcal{Q}^{(v)}_{t}(\{y_{s}\})$ 
%to show that the heat $\mathcal{Q}^{(v)}_{t}(\{y_{s}\})$ 
%is a functional of $\{y_{s}\}$ with $s\in[t_{0},t]$.) 
   On the other hand, the work $\mathcal{W}_{t}^{(v)}(\{y_{s}\})$ 
done on the system to sustain it in a steady state is given by  
\begin{eqnarray}
   \mathcal{W}_{t}^{(v)}(\{y_{s}\}) \equiv \int_{t_{0}}^{t}ds\; 
   \dot{\mathcal{W}}^{(v)}(y_{s}) .
\label{Work1}\end{eqnarray}
using the work rate (\ref{WorkRate1}). 
   The heat (\ref{Heat1}) and the work (\ref{Work1}) 
are related by  
\begin{eqnarray}
   \mathcal{Q}^{(v)}_{t}(\{y_{s}\}) = 
   \mathcal{W}_{t}^{(v)}(\{y_{s}\}) 
   - \Delta \mathcal{U}(y_{t},y_{0})
\label{EnergyBalan1}\end{eqnarray}
with the internal (potential) energy difference 
\begin{eqnarray}
    \Delta \mathcal{U}(y_{t},y_{0}) \equiv 
    U(y_{t}) - U(y_{0}) 
\label{EnergDiffe1}\end{eqnarray}
at times $t$ and $t_{0}$.
   The relation (\ref{EnergyBalan1}) is nothing but the 
energy conservation law satisfied 
even by fluctuating quantities.  
   It may be noted that Eq. (\ref{EnergyBalan1}) is used 
as a ``definition'' of heat in Ref. \cite{ZC03a,ZC04}, 
while here it appears as a consequence 
of our nonequilibrium Onsager-Machlup theory.  
   In other words, our generalization of the Onsager-Machlup theory 
gives a justification of the heat used in Ref. \cite{ZC03a,ZC04}. 
   For other attempts to justify the energy conservation law 
in stochastic processes using a Langevin equation 
or a master equation, see Refs. \cite{S98,C00}.

%---------------------------------------------------------------------
\subsection{Dissipation functions and the entropy production}
\label{DissipationFunctions}

   First, it follows from Eqs. (\ref{DissiFunctA1}) 
and (\ref{EntroProdu1}) that 
\begin{eqnarray}
   \Phi^{(-v)} (-\dot{y}_{s}) &=& \Phi^{(v)} (\dot{y}_{s}) , \\
   \dot{S}^{(-v)}(-\dot{y}_{s},y_{s})
      &=& - \dot{S}^{(v)}(\dot{y}_{s},y_{s}) , 
%\label{}
\end{eqnarray}
implying that the dissipation function 
$\Phi^{(v)} (\dot{y}_{s})$ [as well as $\Psi(y_{s})$ 
by Eq. (\ref{DissiFunctB1})] 
is invariant under the time-reversal changes 
$\dot{y}_{s}\rightarrow -\dot{y}_{s}$ and 
$v\rightarrow -v$, while the 
entropy production rate $\dot{S}^{(v)}(\dot{y}_{s},y_{s})$ 
is anti-symmetric under these changes. 
   It is also obvious from Eqs. (\ref{DissiFunctA1}) 
and (\ref{DissiFunctB1}) that 
\begin{eqnarray}
   \Phi^{(v)} (\dot{y}_{s}) &\geq& 0 , 
      \label{DissiFunctPosit1}\\
   \Psi(y_{s}) &\geq& 0 ,
      \label{DissiFunctPosit2}
\end{eqnarray}
namely, that the dissipation functions are non-negative. 
   One should also notice that 
by the definitions (\ref{DissiFunctA1}) 
and (\ref{DissiFunctB1}) the dissipation functions 
$\Phi^{(v)} (\dot{y}_{s})$ and $\Psi(y_{s})$ 
are proportional to the friction constant $\alpha$. 

   Second, from Eqs. (\ref{RandomForce1}) and (\ref{LangeEq2}), 
the ensemble average $\langle y_{s}\rangle$ of the particle 
position $y_{s}$ satisfies 
\begin{eqnarray}
   \langle \dot{y}_{s} \rangle 
   = -\frac{1}{\tau} \langle y_{s} \rangle - v , 
\label{LangeEq2Ave}\end{eqnarray}
with the time-derivative $\dot{y}_{s} \equiv 
dy_{s}/ds$ of $y_{s}$, 
%so that $\langle y_{s}\rangle$ is given by 
leading to $\langle y_{s} \rangle 
= -v\tau + (\langle y_{0} \rangle +v\tau) 
\exp[-(s-t_{0})/\tau]$. 
   Using this average position  
$\langle y_{s} \rangle$ and 
the average velocity $\langle \dot{y}_{s} \rangle$,   
it follows from Eqs. 
(\ref{DissiFunctA1}), (\ref{DissiFunctB1})  
and (\ref{LangeEq2Ave}) that  
\begin{eqnarray}
   \Phi^{(v)} (\langle \dot{y}_{s} \rangle) 
   = \Psi(\langle y_{s} \rangle) .
\label{DissiFunctRelat1}\end{eqnarray}
   Namely, the two dissipation functions 
$\Phi^{(v)} (\dot{y}_{s})$ and $\Psi(y_{s})$ have the same value 
for $\langle y_{s} \rangle$ and $\langle \dot{y}_{s} \rangle$, 
although $\Phi^{(v)} (\dot{y}_{s})$ is a function of 
$\dot{y}_{s}$ and $\Psi(y_{s})$ is a function of $y_{s}$. 
   Moreover, from Eqs. (\ref{DissiFunctA1}), (\ref{EntroProdu1}), 
(\ref{DissiFunctPosit1}),  (\ref{LangeEq2Ave}) 
and (\ref{DissiFunctRelat1}) we derive   
\begin{eqnarray}
   \dot{S}^{(v)}(\langle \dot{y}_{s} \rangle,
   \langle y_{s} \rangle) 
   = 2\Phi^{(v)} (\langle \dot{y}_{s} \rangle) 
   = 2\Psi (\langle y_{s} \rangle) \geq 0 , 
\label{SeconLaw1}\end{eqnarray}
namely, the function $2\Phi^{(v)} (\langle \dot{y}_{s} \rangle)$ 
[as well as $2\Psi(\langle y_{s} \rangle)$] gives the entropy production rate 
$\dot{S}^{(v)}(\langle \dot{y}_{s} \rangle,\langle y_{s} \rangle)$, 
%for $\langle y_{s} \rangle$ and $\langle \dot{y}_{s} \rangle$, 
justifying the name ``dissipation function'' 
for  $\Phi^{(v)} (\dot{y}_{s})$ and $\Psi(y_{s})$.
   The inequality in (\ref{SeconLaw1}) is 
the second law of thermodynamics in the Onsager-Machlup theory.

%---------------------------------------------------------------------
\subsection{Onsager's principle of minimum energy dissipation and the most probable path}
\label{OnsagerPrincipleMostProbablePath}

   Equation (\ref{LangeEq2Ave}) for the average 
$\langle y_{s}\rangle$ of the particle position 
can be derived from the variational principle 
\begin{eqnarray} 
   \Phi^{(v)} (\dot{y}_{s}) + \Psi(y_{s}) 
   - \dot{\mathcal{S}}^{(v)}(\dot{y}_{s},y_{s}) = 
   \mbox{minimum} , 
\label{VariaPrinc0}\end{eqnarray}
without using the Langevin equation (\ref{LangeEq2}). 
   This can be proved by using that 
$\Phi^{(v)} (\dot{y}_{s}) + \Psi(y_{s}) 
   - \dot{\mathcal{S}}^{(v)}(\dot{y}_{s},y_{s}) 
= -2k_{B}L^{(v)}(\dot{y}_{s},y_{s}) \geq 0$ 
and $L^{(v)}(\langle \dot{y}_{s} \rangle,\langle y_{s} \rangle) 
=0$, so that the left-hand side of Eq.  (\ref{VariaPrinc0}) 
takes its minimum value for $y_{s} = \langle y_{s} \rangle$, 
i.e. for the \textit{average} path, which is used in 
Eq. (\ref{SeconLaw1}). 
   Equation (\ref{VariaPrinc0}) is called 
the Onsager's principle of minimum energy 
dissipation, and is proposed as a generalization 
of the maximal entropy principle for equilibrium 
thermodynamics \cite{OM53,MO53,TM57,O31a,O31b,H76,H77}. 

   Another result in the Onsager-Machlup theory as 
a variational principle is that 
we can justify a variational principle to extract the 
special path $\{y_{s}^{*}\}_{s\in[t_{0},t]}$, 
the so-called \textit{most probable path}, which 
give the most significant contribution in the transition 
probability $\transS{F}{y_{t}}{t}{y_{0}}{t_{0}}$. 
   By the expression (\ref{TransProba1}) for the transition 
probability $\transS{F}{y_{t}}{t}{y_{0}}{t_{0}}$, 
the most probable path $\{y_{s}^{*}\}_{s\in[t_{0},t]}$ is 
determined by the maximal condition on  
$\int_{t_{0}}^{t}ds\; L^{(v)}\!\left(\dot{y}_{s},y_{s}\right)$, 
in other words, the path $\{y_{s}\}_{s\in[t_{0},t]}$ 
satisfying 
\begin{eqnarray} 
   \int_{t_{0}}^{t}ds\; \left[\Phi^{(v)} (\dot{y}_{s}) + \Psi(y_{s}) 
   - \dot{\mathcal{S}}^{(v)}(\dot{y}_{s},y_{s})\right] = 
   \mbox{minimum} , 
   \nonumber \\ 
\label{VariaPrinc1}\end{eqnarray}
under fixed values of $y_{0}$ and $y_{t}$, 
noting the expression (\ref{Lagra2}) for 
the Lagrangian function $L^{(v)}\!\left(\dot{y}_{s},y_{s}\right)$. 
%
%   We can discuss the same principle as Eq. (\ref{VariaPrinc1}) 
% in another way 
% using the Lagrangian 
% function $L^{(v)}\!\left(\dot{y}_{s},y_{s}\right)$ itself. 
   The condition (\ref{VariaPrinc1}), or equivalently 
the maximal condition of  
$\int_{t_{0}}^{t}ds\; L^{(v)}\!\left(\dot{y}_{s},y_{s}\right)$ 
implies the variational principle 
$\delta \int_{t_{0}}^{t}ds\; L^{(v)}\!\left(\dot{y}_{s},y_{s}\right) 
= 0$ for the path $\{y_{s}\}_{s\in[t_{0},t]}$, leading to 
the Euler-Lagrange equation 
\cite{LL69} 
\begin{eqnarray} 
   \frac{d}{ds} 
   \frac{\partial L^{(v)}(\dot{y}^{*}_{s},y^{*}_{s})}
      {\partial \dot{y}^{*}_{s}}
   - \frac{\partial L^{(v)}(\dot{y}^{*}_{s},y^{*}_{s})}
      {\partial y^{*}_{s}} = 0 
\label{EularLagra1}\end{eqnarray}
for the most probable path $\{y_{s}^{*}\}_{s\in[t_{0},t]}$. 
   (The most probable path can also be analyzed by 
the Hamilton-Jacobi equation \cite{BSG01,BSG02}.)
   Inserting Eq. (\ref{Lagra1}) into Eq. (\ref{EularLagra1}) 
we obtain 
\begin{eqnarray} 
   \frac{d^{2} y^{*}_{s}}{ds^{2}} 
      = \frac{y^{*}_{s} + v\tau}{\tau^{2}} 
\label{EularLagra2}\end{eqnarray}
for our model. 
   It is interesting to note that the ensemble average 
$\langle y_{s}\rangle$ also satisfies Eq.  
(\ref{EularLagra2}), because  
$d^{2}\langle y_{s}\rangle/ds^{2} 
= (\langle y_{s}\rangle + v\tau)/\tau^{2}$ from 
Eq. (\ref{LangeEq2Ave}). 
   In general, the most probable path 
$\{y_{s}^{*}\}_{s\in[t_{0},t]}$ with the conditions 
$y_{t_{0}}^{*}=y_{0}$ and $y_{t}^{*}=y_{t}$ contains  
a superposition of the forward average path 
%$\{\langle y_{t}\rangle\}_{s\in[t_{0},t]}$ 
$\Upsilon_{s}^{[+]} \equiv \mathcal{A}_{+} \exp(-s/\tau) -v\tau$ 
(like the average path $\langle y_{t}\rangle$) 
and its time-reversed path 
$\Upsilon_{s}^{[-]} \equiv \mathcal{A}_{-} \exp(s/\tau) +v\tau$, namely  
\begin{eqnarray}
   y^{*}_{s} = \Upsilon_{s}^{[+]} + \Upsilon_{s}^{[-]} 
          + \mathcal{A}_{0}
%\label{}
\end{eqnarray}
where $\mathcal{A}_{\pm}$ and $\mathcal{A}_{0} (=-v\tau)$  
are time-independent constants and are determined by the 
conditions 
$y_{t_{0}}^{*}=y_{0}$ and $y_{t}^{*}=y_{t}$ \cite{noteIIIC}. 
 
   We now discuss a relation of the Onsager-Machlup theory 
with Einstein's fluctuation formula \cite{Lan59}.  
   We note that   
\begin{eqnarray}
   L^{(v)}\! \left(\dot{\Upsilon}_{s}^{[+]},
      \Upsilon_{s}^{[+]}\right) &=& 0, 
      \label{LagraForwaPath1}\\
   L^{(v)}\! \left(\dot{\Upsilon}_{s}^{[-]},
   \Upsilon_{s}^{[-]}\right) &=& 
   \frac{1}{k_{B}}   
   \dot{S}^{(v)}\!\left(\dot{\Upsilon}_{s}^{[-]},
   \Upsilon_{s}^{[-]}\right)
   \label{LagraBackPath1}
\end{eqnarray}
with $\dot{\Upsilon}_{s}^{[\pm]}\equiv d \Upsilon_{s}^{[\pm]}/ds$. 
   Here, we used the equation $\pm d \Upsilon_{s}^{[\pm]}/ds 
= -\Upsilon_{s}^{[\pm]}/\tau \mp v$. 
%   Note that by Eqs. (\ref{Lagra2}), (\ref{LagraForwaPath1}) 
%and $\Phi^{(v)} (\dot{y}_{s}) + \Psi(y_{s}) 
%- \dot{\mathcal{S}}^{(v)}(\dot{y}_{s},y_{s})\geq 0$ 
%the forward path $\Upsilon_{s}^{[+]}$ makes the left-hand 
%side of the variational principle (\ref{VariaPrinc1}) 
%minimum without its time-integral. 
%  
   Using the most probable path 
$\{y_{s}^{*}\}_{s\in[t_{0},t]}$ satisfying the conditions 
$y_{t_{0}}^{*}=y_{0}$ and $y_{t}^{*}=y_{t}$, we can approximate 
the transition probability $\transS{F}{y_{t}}{t}{y_{0}}{t_{0}}$ as  
\begin{eqnarray}
   \transP{F}{y_{t}}{t}{y_{0}}{t_{0}} 
   \approx \exp\left[
      \int_{t_{0}}^{t}ds\; L^{(v)}\!
      \left(\dot{y}^{*}_{s},y^{*}_{s}\right)\right] ,
%   \nonumber
\label{OMformu1}\end{eqnarray}
apart from a normalization factor. 
   This is analogous to the classical approximation for  
the wave function in the Feynman path-integral approach 
in quantum mechanics \cite{FH65}. 
   It is meaningful to mention that in the case 
of relaxation to an equilibrium  state  
($v=0$),  Eq. (\ref{OMformu1}) becomes 
\begin{eqnarray}
   &&\left.\transP{F}{y_{t}}{t}{y_{0}}{t_{0}}\right|_{v=0} 
   \nonumber \\
   &&\spaEq\approx \exp\left[
      \frac{1}{k_{B}}\int_{t_{0}}^{t}ds\; \left.
      \dot{S}^{(v)}\!\left(\dot{\Upsilon}_{s}^{[-]},
      \Upsilon_{s}^{[-]}\right)\right|_{v=0}\right] , \;\;\;
\label{OMformu2}\end{eqnarray}
noting Eqs. (\ref{LagraBackPath1}) and 
$L^{(v)}(\dot{y}^{*}_{s},y^{*}_{s})|_{v=0}
=L^{(v)}(\dot{\Upsilon}_{s}^{[-]},\Upsilon_{s}^{[-]})|_{v=0}$ 
\cite{TM57}. 
   Here, we remark that in Eq. (\ref{OMformu2}) 
the transition probability 
$\transS{F}{y_{t}}{t}{y_{0}}{t_{0}}|_{v=0}$ is expressed 
by the time-reversed path 
$\{\dot{\Upsilon}_{s}^{[-]}\}_{s\in[t_{0},t]}$ only. 
   The quantity $\int_{t_{0}}^{t}ds\; 
\dot{S}$ gives the entropy, so that Eq. (\ref{OMformu2}) 
corresponds to Einstein's fluctuation formula 
in equilibrium, i.e. for $v=0$.

%%%%%%%%%%%%%%%%%%%%%%%%%%%%%%%%%%%%%%%%%%%%%%%%%%%%%%%%%%%%%%%%%%%%%%
\section{Fluctuation Theorem for Work}
\label{FluctuationTheoremWork}

   In the preceding section \ref{OnsagerMachlupSteadyStates},  
by generalizing the Onsager-Machlup theory to   
nonequilibrium steady states,  
we discussed fluctuating quantities whose averages give 
thermodynamic quantities, like work and heat, 
etc. 
   Since these quantities fluctuate, it is important 
to discuss nonequilibrium characteristics of their fluctuations. 
   In the remaining part of this paper, we discuss such  
characteristics using distribution functions of work, heat, etc., 
by the functional integral technique. 
   For this discussion, generalized versions of the equilibrium 
detailed balance, which we will call nonequilibrium detailed 
balance relations, play an important role, 
leading to fluctuation theorems. 
%\cite{ZC03a,ECM93,GC95,G96,ES02}. 
   Fluctuation theorems are for nonequilibrium 
behavior in the case of $v\neq 0$, 
so there is no counterpart to the contents of this paper 
in Onsager and Machlup's original papers 
%\cite{OM53,MO53} 
where $v=0$ always. 
%   It may be noted that the work $\mathcal{W}_{t}^{(v)}(\{y_{s}\})$, 
%the heat $\mathcal{Q}^{(v)}_{t}(\{y_{s}\})$, etc., are introduced as 
%functionals involving a time-integral of a 
%function defined at each time, which is called 
%a Wiener functional \cite{W23,K49}.  
%   The functional integral technique 
%can be an especially powerful tool to discuss the distributions 
%of such functionals. 

%---------------------------------------------------------------------
\subsection{Nonequilibrium detailed balance relation}

   The equilibrium detailed balance condition expresses 
a reversibility of the transition probability between any 
two states in the equilibrium state, and is known as 
a physical condition for the system to relax to 
an equilibrium state \cite{K92,R89}.  
   This condition has to be modified 
for the nonequilibrium steady state, 
because the system does not relax to an equilibrium state 
but is sustained in an nonequilibrium state 
by an external force. 
%\cite{K98,LS99}. 
   This modification, or violation, of the equilibrium 
detailed balance in the nonequilibrium steady state is 
expressed quantitatively for work by 
\begin{eqnarray}
   && 
   e^{-\beta\mathcal{W}_{t}^{(v)}(\{y_{s}\})}
      e^{\int_{t_{0}}^{t}ds\; L^{(v)}\!
      \left(\dot{y}_{s},y_{s}\right)}
      f_{eq}(y_{0}) 
      \nonumber \\
   && \spaEq 
   = f_{eq}(y_{t}) 
      \;e^{\int_{t_{0}}^{t}ds\; L^{(-v)}\!
      \left(-\dot{y}_{s},y_{s}\right)}
\label{DetaiBalan1}\end{eqnarray}
in our path-integral approach, which is derived from Eqs. 
(\ref{Lagra1}), (\ref{EquilDistr1}) and (\ref{Work1}).  
   We call Eq. (\ref{DetaiBalan1}) 
a \emph{nonequilibrium detailed balance relation} 
for nonequilibrium steady states in this paper \cite{noteIVa}. 
   Equation (\ref{DetaiBalan1}) reduces 
to the equilibrium detailed balance condition  
in the case $v=0$, because  
from Eqs. (\ref{TransProba1}), (\ref{DetaiBalan1}) 
and $\mathcal{W}_{t}^{(0)}(\{y_{s}\})$ $=0$, we can derive 
the well-known equilibrium detailed balance condition  
\begin{eqnarray}
   \left.\transP{F}{y_{t}}{t}{y_{0}}{t_{0}}\right|_{v=0} 
   f_{eq}(y_{0}) 
   = \left.\transP{F}{y_{0}}{t}{y_{t}}{t_{0}}\right|_{v=0} 
   f_{eq}(y_{t})
\label{DetaiBalan2}\end{eqnarray}
%
%in the case of $v=0$, 
for the transition probability $\transS{F}{y_{t}}{t}{y_{0}}{t_{0}}$ 
in equilibrium. 

   As discussed in Sec. \ref{DraggedBrownianParticle}, 
the term $\exp[\int_{t_{0}}^{t}ds\; L^{(v)}(\dot{y}_{s},y_{s})]$ 
on the left-hand side of Eq. (\ref{DetaiBalan1}) 
is the probability functional for the forward path 
$\{y_{s}\}_{s\in[t_{0},t]}$. 
   On the other hand, the term 
$\exp[\int_{t_{0}}^{t}ds\; L^{(-v)}(-\dot{y}_{s},y_{s})]$ 
on the right-hand side of Eq. (\ref{DetaiBalan1}) is 
the probability functional of the time-reversed path.  
   Therefore,  Eq. (\ref{DetaiBalan1}) means that 
we need the work $\mathcal{W}_{t}^{(v)}(\{y_{s}\})$ 
so that the particle, dragged from an equilibrium state 
with the velocity $v$,  
can move along a path $\{y_{s}\}_{s\in[t_{0},t]}$ 
and return back to the equilibrium state along 
its time-reversed path with the reversed dragging velocity $-v$. 
   Such an additional work appears as a canonical 
distribution type of barrier    
$\exp[-\beta\mathcal{W}_{t}^{(v)}(\{y_{s}\})]$ 
for the transition probability  
on the left-hand side of Eq. (\ref{DetaiBalan1}). 
   It should be emphasized that Eq. (\ref{DetaiBalan1}) is 
satisfied not only for the most probable path but  
for \emph{any} path $\{y_{s}\}_{s\in[t_{0},t]}$, 
which is crucial for the derivation of the work fluctuation theorem 
as we will discuss in the next subsection 
\ref{WorkFluctuationTheorem}.

%---------------------------------------------------------------------
\subsection{Work fluctuation theorem}
\label{WorkFluctuationTheorem}

   Now, we discuss the distribution of work. 
   For simplicity of notation, we consider the 
dimensionless work $\beta\mathcal{W}_{t}^{(v)}(\{y_{s}\})$  
and its distribution $P_{w}(W,t)$ given by 
\begin{eqnarray}
   P_{w}(W,t) = \pathaveA{\delta\!\left(W - 
   \beta\mathcal{W}_{t}^{(v)}(\{y_{s}\})\right)} .
\label{WorkDistr1}\end{eqnarray}
   Here, $\pathaveA{\cdots}$ means a functional average
over all possible paths $\{y_{s}\}_{s\in[t_{0},t]}$,  
as well as integrals over the initial and final points of the path:    
\begin{eqnarray}
   \pathaveA{\; X_{t}(\{y_{s}\})\;} 
   &\equiv&
  %\equiv 
      \int dy_{t} \! \int_{y_{0}}^{y_{t}} 
      \mathcal{D}y_{s} \!\int dy_{0} \;   
      e^{\int_{t_{0}}^{t}ds\; 
      L^{(v)}\!\left(\dot{y}_{s},y_{s}\right)}
      \nonumber \\
   &&\spaEq \times 
   \; f(y_{0},t_{0}) \; X_{t}(\{y_{s}\}) 
\label{PathAvera1}\end{eqnarray}
for any functional $X_{t}(\{y_{s}\})$. 
   It is convenient to express the work distribution 
$P_{w}(W,t)$ as a Fourier transform 
\begin{eqnarray}
   P_{w}(W,t) = \frac{1}{2\pi}\int_{-\infty}^{+\infty}d\lambda \;
   e^{i\lambda W} \mathcal{E}_{w}^{(v)}(i\lambda,t)
\label{WorkDistr2}\end{eqnarray}
using the function $\mathcal{E}_{w}^{(v)}(\lambda,t)$ defined by 
\begin{eqnarray}
   \mathcal{E}_{w}^{(v)}(\lambda,t) \equiv 
   \pathaveA{e^{-\lambda\beta 
   \mathcal{W}_{t}^{(v)}(\{y_{s}\})}} ,
\label{EFunctWork1}\end{eqnarray}
which may be regarded as a generating functional 
of the dimensionless work. 
   It follows from 
%Eqs. (\ref{Lagra1}), 
Eqs. (\ref{PathAvera1}), (\ref{EFunctWork1}), 
$L^{(-v)}(\dot{y}_{s},y_{s}) 
= L^{(v)}(-\dot{y}_{s},-y_{s})$ 
and $\mathcal{W}_{t}^{(-v)}(\{y_{s}\}) 
=\mathcal{W}_{t}^{(v)}(\{-y_{s}\})$, 
that the function 
$\mathcal{E}_{w}^{(v)}(\lambda,t)$ 
is invariant under the change $v\rightarrow -v$, 
namely  
\begin{eqnarray}
   \mathcal{E}_{w}^{(-v)}(\lambda,t) 
   = \mathcal{E}_{w}^{(v)}(\lambda,t),  
\label{EFunctWorkRelat1}\end{eqnarray}
if the initial distribution $f(y_{0},t_{0})$ is invariant 
under 
%the changes $y_{0}\rightarrow -y_{0}$ 
%[and $v\rightarrow -v$, e.g. for the initial distribution 
%of Eq. (\ref{SteadSolut1})], 
%namely, 
spatial reflection, namely 
$f(-y_{0},t_{0})|_{-v}=f(y_{0},t_{0})|_{v}$.
   This is simply due to an invariance under space inversion 
of our model. 

   In addition, as shown in Appendix 
\ref{TransientFluctuationTheoremWork},  
the nonequilibrium detailed balance 
relation (\ref{DetaiBalan1}) imposes the relation 
$\mathcal{E}_{w}^{(v)}(\lambda,t) 
= \mathcal{E}_{w}^{(-v)}(1-\lambda,t)$ 
on the function $\mathcal{E}_{w}^{(v)}(\lambda,t)$. 
   Combination of this relation  
with Eq. (\ref{EFunctWorkRelat1}) then leads to 
\begin{eqnarray}
    \mathcal{E}_{w}^{(v)}(\lambda,t) 
    = \mathcal{E}_{w}^{(v)}(1-\lambda,t)
\label{FluctTheorWork1}\end{eqnarray}
for the equilibrium initial distribution  $f(y_{0},t_{0}) 
=f_{eq}(y_{0})$, as a form discussed in Ref. \cite{LS99}.  
   Equation (\ref{FluctTheorWork1}) is equivalent to the relation 
\begin{eqnarray}
   \frac{P_{w}(W,t)}{P_{w}(-W,t)} = \exp(W).
\label{FluctTheorWork2}\end{eqnarray}
for the work distribution $P_{w}(W,t)$, which is known 
as the transient fluctuation theorem \cite{ZC03b,ES02,noteIIIB}. 
   [See Appendix \ref{TransientFluctuationTheoremWork} 
for a derivation of Eq. (\ref{FluctTheorWork2}) from 
 Eq. (\ref{FluctTheorWork1}).]

   As shown in Eq. (\ref{FluctTheorWork2}),  
the transient fluctuation theorem is satisfied for any time 
being an identity \cite{CG99},  
but it requires that the system is in the equilibrium state at the 
initial time $t_{0}$. 
%   The fluctuation theorem under such a restriction of the 
%initial distribution is called the transient fluctuation theorem 
%\cite{ES02}. 
   Therefore, one may ask what happens 
to the fluctuation theorem if we 
choose a nonequilibrium steady state, or any other 
state, as the initial condition. 
   In the next subsection 
\ref{FunctionalCalculationWorkDistribution}, 
we calculate the work distribution function 
$P_{w}(W,t)$ explicitly by carrying out the functional 
integral on the right-hand side of Eq. (\ref{WorkDistr1}) 
via Eq. (\ref{PathAvera1}), 
in order to answer this question.

%---------------------------------------------------------------------
\subsection{Functional integral calculation of the work distribution function}
\label{FunctionalCalculationWorkDistribution}

   To calculate the work distribution function $P_{w}(W,t)$, 
we note first that the function $\mathcal{E}_{w}^{(v)}(\lambda,t)$,  
connected to $P_{w}(W,t)$ by Eq. (\ref{WorkDistr2}), can be  
rewritten as  
\begin{eqnarray}
   \mathcal{E}_{w}^{(v)}(\lambda,t) 
   = \int dy_{t} \! \int dy_{0} \; \mathcal{F}(y_{t},y_{0};\lambda)  
      f(y_{0},t_{0}) 
\label{EFunctWork2}\end{eqnarray}
by Eqs. (\ref{PathAvera1}) and (\ref{EFunctWork1}).  
   Here, $\mathcal{F}(y_{t},y_{0};\lambda)$ is defined by 
\begin{eqnarray}
   &&\!\!\!\mathcal{F}(y_{t},y_{0};\lambda) 
   \nonumber \\
  && \equiv
%   \equiv 
      \int_{y_{0}}^{y_{t}} \mathcal{D}y_{s} \; 
%      \nonumber \\
%    &&\times 
       \exp\left\{\int_{t_{0}}^{t}ds\; \left[
      L^{(v)}\!\left(\dot{y}_{s},y_{s}\right)
      -\lambda\beta\dot{\mathcal{W}}^{(v)}(y_{s})\right]\right\} .
      \nonumber \\
\label{EFunctWorkTrnas1}\end{eqnarray}
   Equation (\ref{EFunctWorkTrnas1}) may be regarded 
as a constrained transition probability 
for the modified Lagrangian 
$L^{(v)}\!\left(\dot{y}_{s},y_{s}\right) 
-\lambda\dot{\mathcal{W}}^{(v)}(y_{s})$ 
\cite{noteIVc}. 
   Here, the $v$-dependence of the function 
$\mathcal{F}(y_{t},y_{0};\lambda)$ has been suppressed, 
as it is in the rest of the paper.  

   To calculate the function $\mathcal{F}(y_{t},y_{0};\lambda)$, 
we introduce the solution $\ym_{t}$ of the modified 
Euler-Lagrange equation 
for the modified Lagrangian $L^{(v)}\!\left(\dot{y}_{s,}y_{s}\right)
-\lambda\dot{\mathcal{W}}^{(v)}(y_{s})$, namely  
\begin{eqnarray} 
   \frac{d}{ds} 
   \frac{\partial  L^{(v)}(\ymd_{s}, \ym_{s})}
      {\partial \ymd_{s}}
   - \frac{\partial L^{(v)}(\ymd_{s},\ym_{s})}
      {\partial \ym_{s}} 
   + \lambda\beta\frac{\partial  \dot{\mathcal{W}}^{(v)}(\ym_{s})}
      {\partial \ym_{s}} = 0 
   \nonumber \\
\label{EularLagra2Modif1}\end{eqnarray}
under the conditions $\ym_{t}=y_{t}$ and 
$(\ym_{0}\equiv)\ym_{t_{0}}=y_{0}$. 
   By solving Eq. (\ref{EularLagra2Modif1}) we obtain 
\begin{eqnarray} 
   \ym_{s} 
   &=& -(1-2\lambda)v\tau + 
      A_{t-t_{0}}^{((1-2\lambda)v)}(y_{t},y_{0})
      \exp\left(-\frac{t-s}{\tau}\right) 
      \nonumber \\
   &&\spaEq 
      + A_{t-t_{0}}^{((1-2\lambda)v)}(y_{0},y_{t})
      \exp\left(-\frac{s-t_{0}}{\tau}\right)
\label{EularLagra2Solve1}\end{eqnarray}
where $A_{t-t_{0}}^{(v)}(y_{t},y_{0})$ is defined by
\begin{eqnarray}
   A_{t-t_{0}}^{(v)}(y_{t},y_{0})
      \equiv\frac{(y_{t} +v\tau) - (y_{0} +v\tau)
      b_{t} }
      {1-b_{t}^{2}} .
\label{ConstA1} \end{eqnarray}
   [See Appendix \ref{CalculationWorkDistribution} 
for a derivation of Eq. (\ref{EularLagra2Solve1}).]
   The path $\{\ym_{s}\}_{s\in[t_{0},t]}$ becomes 
the most probable path $\{y_{s}^{*}\}_{s\in[t_{0},t]}$ 
in the case of $\lambda=0$ 
in which Eq. (\ref{EularLagra2Modif1}) is 
equivalent to Eq. (\ref{EularLagra1}).
% which was discussed in Sec. 
%\ref{OnsagerPrincipleMostProbablePath}. 

   Using the solution $\ym_{s}$ of the modified Euler-Lagrange 
equation (\ref{EularLagra2Solve1}), we obtain  
\begin{eqnarray}
   \mathcal{F}(y_{t},y_{0};\lambda)
   &=& 
%   = 
   e^{\int_{t_{0}}^{t} ds 
      \left[L^{(v)}(\dot{\ym}_{s},\ym_{s})
      -\lambda \beta\dot{\mathcal{W}}^{(v)}(\ym_{s})\right]}
      \nonumber \\
   &&\times
      \int_{\tilde{z}_{0}}^{\tilde{z}_{t}} 
      \mathcal{D}\tilde{z}_{s} \;   
      e^{\int_{t_{0}}^{t} ds \; 
      L^{(0)}(\dot{\tilde{z}}_{s},\tilde{z}_{s})}
%      \nonumber \\
\label{EFunctWorkTrnas2}\end{eqnarray}
for the function $\mathcal{F}(y_{t},y_{0};\lambda)$, 
where $\tilde{z}_{s}$ is introduced as 
\begin{eqnarray}
   \tilde{z}_{s} \equiv y_{s} - \ym_{s}, 
\label{VariaZ1}\end{eqnarray}
namely the deviation of $y_{s}$ from $\ym_{s}$,  
satisfying the boundary conditions 
\begin{eqnarray}
   \tilde{z}_{t} = \tilde{z}_{0} = 0 
\label{VariaZBound1}\end{eqnarray}
%. 
because $\ym_{0}=y_{0}$ and $\ym_{t}=y_{t}$, where  
$\dot{\tilde{z}}_{s}\equiv d\tilde{z}_{s}/ds$ 
and $\tilde{z}_{0}\equiv \tilde{z}_{t_{0}}$.
   [See Appendix \ref{CalculationWorkDistribution} for a 
derivation of Eq. (\ref{EFunctWorkTrnas2}).]
   For the functional integral for $\tilde{z}_{s}$ 
on the right-hand side of Eq. (\ref{EFunctWorkTrnas2}) 
we obtain  
\begin{eqnarray}
%c   &&
   \int_{\tilde{z}_{0}}^{\tilde{z}_{t}} 
      \mathcal{D}\tilde{z}_{s} \;   
      \exp\left[\int_{t_{0}}^{t} ds \; 
      L^{(0)}(\dot{\tilde{z}}_{s},\tilde{z}_{s})\right]
      =\frac{1}{\sqrt{4\pi D \mathcal{T}_{t}}} ,
%c      \nonumber \\
\label{PathInteg1}\end{eqnarray}
noting that the Lagrangian 
$L^{(0)}(\dot{\tilde{z}}_{s},\tilde{z}_{s})$ on the left-hand 
side of Eq. (\ref{PathInteg1}) is for the case of $v=0$. 
   [See Appendix \ref{CalculationWorkDistribution} for a 
derivation of Eq. (\ref{PathInteg1}).]
   Inserting Eqs. (\ref{EularLagra2Solve1}) and (\ref{PathInteg1}) 
into Eq. (\ref{EFunctWorkTrnas2}), 
the function $\mathcal{F}(y_{t},y_{t_{0}};\lambda)$ 
is represented as 
\begin{widetext}
\begin{eqnarray}
   \mathcal{F}(y,y_{0};\lambda)
      &=& \frac{1}{\sqrt{4\pi D 
      \mathcal{T}_{t}}}
      \exp\left\{- \frac{\left[(y +v\tau) - (y_{0} +v\tau)
      b_{t} \right]^{2}}
      {4D\mathcal{T}_{t}}
%      \right.\nonumber \\
%   && \spaEq \left.
      + \lambda \alpha \beta v \left(y + y_{0}\right)
      \frac{1-b_{t}}
      {1+b_{t}}
      \right.\nonumber \\
   && \spaEq\spaEq\spaEq \left.
      - \lambda (1-\lambda) \alpha \beta v^{2} \left( t-t_{0}
      - 2\tau \frac{1-b_{t}}
      {1+b_{t}} \right)
      \right\} 
%      \nonumber \\
\label{FunctCalF1}\end{eqnarray}
   Using Eq. (\ref{EFunctWork2})  and (\ref{FunctCalF1}) 
and carrying out the integration over $y$ 
we obtain 
\begin{eqnarray}
   \mathcal{E}_{w}^{(v)}(\lambda,t) 
%c   &=& 
   =
   e^{ - \lambda (1-\lambda) \alpha \beta v \Omega_{t}} 
%c      \nonumber \\
%c   &&\times 
      \int dy_{0}\; f\!\left(y_{0},t_{0}\right) 
%c      \nonumber \\
%c   &&\times
      e^{ \lambda \alpha\beta v 
     \left[y_{0} - \frac{v\tau}{2} 
      \left(1-b_{t}\right)\right]
      \left(1 - b_{t}\right)}
\label{EFunctWork3}\end{eqnarray}
where $\Omega_{t}$ is defined by 
\begin{eqnarray}
    \Omega_{t}
    \equiv v \left\{ t-t_{0}
      - \frac{\tau}{2} \left[ 4 
      - \left(1-b_{t}\right)^{2}
      \right]\frac{1-b_{t}}
      {1+b_{t}} \right\} .
%c      \nonumber \\
\label{FunctOmega1}\end{eqnarray}
   Equation (\ref{EFunctWork3}) gives an explicit 
expression of the function $\mathcal{E}_{w}^{(v)}(\lambda,t)$ 
for any initial distribution $f(y_{0},t_{0})$. 
%ap    [We give calculation details of Eqs. (\ref{EularLagra2Solve1}), 
%ap (\ref{PathInteg1}) and (\ref{EFunctWork3}) in Appendix 
%ap \ref{CalculationWorkDistribution}.]

   Inserting Eq. (\ref{EFunctWork3}) into 
Eq. (\ref{WorkDistr2}), we obtain 
%
%\begin{widetext}
\begin{eqnarray}
   P_{w}(W,t) 
   = 
   \frac{1}{\sqrt{
      4\pi\alpha\beta v \Omega_{t}}} 
      \int dy_{0}\; f\!\left(y_{0} ,t_{0} \right)
%      \nonumber \\
%   &&\spaEq \times 
      \exp\left\{-\frac{\left\{W-\alpha\beta v 
      \left[
      \Omega_{t} -\left(1-b_{t}\right) 
      \left[y_{0}-\frac{v\tau}{2}
      \left(1-b_{t}\right)\right]
      \right]\right\}^{2}}
      {4\alpha\beta v \Omega_{t}}\right\} 
%    \\
%   &=& 
%   \frac{1}{\sqrt{
%      4\pi\alpha\beta v \Omega_{t}}} 
%      \int dy\; f\!\left(y +\frac{v\tau}{2}
%      \left(1-b_{t}\right),t_{0} \right)
%      \nonumber \\
%   &&\spaEq \times 
%      \exp\left\{-\frac{\left[W-\alpha\beta v \left(
%      \Omega_{t} -\left(1-b_{t}\right) y
%      \right)\right]^{2}}
%      {4\alpha\beta v \Omega_{t}}\right\} .
%c      \nonumber \\
\label{WorkDistr3}\end{eqnarray}
\end{widetext}
as a concrete form of the work distribution $P_{w}(W,t)$ 
satisfied by \emph{any} initial distribution $f(y_{0},t_{0})$. 
   From Eq.  (\ref{WorkDistr3}), the asymptotic form 
of the work distribution function $P_{w}(W,t)$ 
is given by 
\begin{eqnarray}
   P_{w}(W,t) \stackrel{t\rightarrow+\infty}{\sim}
   \frac{1}{\sqrt{4\pi \alpha\beta v^{2} t}}
   \exp\left[-\frac{
   \left(W-\alpha\beta v^{2} t\right)^{2}}
   {4\alpha\beta v^{2} t}
   \right] 
\label{DistrWorkLongTime1}\end{eqnarray}
where we used the asymptotic relation 
$\Omega_{t}\stackrel{t\rightarrow+\infty}{\sim} vt$ 
by Eq. (\ref{FunctOmega1}), 
and the normalization condition 
$\int dy_{0}\; f(y_{0} ,t_{0})=1$.
   It follows immediately 
from Eq. (\ref{DistrWorkLongTime1}) that  
\begin{eqnarray}
   \lim_{t\rightarrow+\infty}
   \frac{P_{w}(W,t)}{P_{w}(-W,t)} = \exp(W) .
\label{FluctTheorWork3}\end{eqnarray}
   Therefore, the work fluctuation theorem is satisfied 
for \textit{any} initial condition 
(including the nonequilibrium steady state 
initial distribution) in the very long time limit.

%%%%%%%%%%%%%%%%%%%%%%%%%%%%%%%%%%%%%%%%%%%%%%%%%%%%%%%%%%%%%%%%%%%%%%
\section{Fluctuation Theorem for Friction} 
\label{FluctuationTheoremFriction}

   In Sec. \ref{FluctuationTheoremWork}, we emphasized 
a close relation between the nonequilibrium 
detailed balance relation like Eq. (\ref{DetaiBalan1}) 
and the work fluctuation theorem (\ref{FluctTheorWork3}). 
   To show the usefulness of such a relation 
we discuss in this section 
another type of nonequilibrium detailed balance relation, 
and show that it leads to another fluctuation theorem 
related to the energy loss by friction. 

   We consider the rate of energy loss caused by 
the friction force $-\alpha \dot{y}_{s}$ in the comoving frame. 
   It is given by $- \alpha \dot{y}_{s} v$, so the total 
energy loss $\mathcal{R}_{t}^{(v)}$ by friction  
in the time interval $[t_{0},t]$ is 
\begin{eqnarray}
   \mathcal{R}_{t}^{(v)}(y_{t},y_{0}) 
   = \int_{t_{0}}^{t}ds\; \left(- \alpha \dot{y}_{s} \right) v 
   = - \alpha v (y_{t}-y_{0}) 
\label{FrictLoss}\end{eqnarray}
using $y_{0}\equiv y_{t_{0}}$.
   It may be noted that the energy loss  
$\mathcal{R}_{t}^{(v)}(y_{t},y_{0})$ by friction is 
determined by the particle positions at the times 
$t_{0}$ and $t$ only, 
different from the 
work $\mathcal{W}_{t}^{(v)}(\{y_{s}\})$, 
which is determined 
by the particle positions at \emph{all} times $s\in[t_{0},t]$.

   Our starting point to discuss the fluctuation theorem 
for friction is the relation 
\begin{eqnarray}
   && 
   e^{-\beta\mathcal{R}_{t}^{(v)}(y_{t},y_{0})}
      e^{\int_{t_{0}}^{t}ds\; L^{(v)}\!
      \left(\dot{y}_{s},y_{s}\right)}
      f_{eq}(y_{0}) 
      \nonumber \\
   && \spaEq 
   = f_{eq}(y_{t}) 
      \;e^{\int_{t_{0}}^{t}ds\; L^{(v)}\!
      \left(-\dot{y}_{s},y_{s}\right)}
\label{DetaiBalan3}\end{eqnarray}
derived straight-forwardly from Eqs. (\ref{Lagra1}), 
(\ref{EquilDistr1}) and (\ref{FrictLoss}).  
  It must be noted that there is a difference 
between Eq. (\ref{DetaiBalan3}) and Eq. (\ref{DetaiBalan1}) 
in the change (or no change) of sign of the dragging 
velocity $v$ in their time-reversed movement on 
their right-hand sides. 
   This difference leads to different fluctuation theorems 
as shown later in this section. 
  Noting this difference, Eq. (\ref{DetaiBalan3}) can be 
interpreted as that the energy loss  
$\mathcal{R}_{t}^{(v)}(y_{t},y_{0})$ by friction is required 
to move the particle from $y_{0}$ 
to $y_{t}$ via the path $\{y_{s}\}_{s\in [t_{0},t]}$ and 
to return it back from $y_{t}$ to $y_{0}$ via its reversed path 
\textit{without} changing the dragging velocity $v$. 
   Using Eqs. (\ref{TransProba1}) and (\ref{DetaiBalan3}) 
we obtain 
\begin{eqnarray}
   e^{-\beta\mathcal{R}_{t}^{(v)}(y_{t},y_{0})}
   \transP{F}{y_{t}}{t}{y_{0}}{t_{0}} f_{eq}(y_{0}) 
   = \transP{F}{y_{0}}{t}{y_{t}}{t_{0}} f_{eq}(y_{t}) ,
   \nonumber \\
\label{DetaiBalan4}\end{eqnarray}
where we used $\transS{F}{y_{0}}{t}{y_{t}}{t_{0}} 
= \int_{y_{0}}^{y_{t}}\mathcal{D}y_{s}\; 
\exp[\int_{t_{0}}^{t}ds\; $ $ L^{(v)}(-\dot{y}_{s},y_{s})]$,  
as shown by Eqs. (\ref{TransProba3}) and (\ref{TransProba4}) 
\cite{RCW04}. 
   Equation (\ref{DetaiBalan4}) reduces to  
the equilibrium detailed balance (\ref{DetaiBalan2}) 
in the case of $v=0$ because of  
$\mathcal{R}_{t}^{(0)}(y_{t},y_{0})=0$. 
   Therefore, Eq. (\ref{DetaiBalan3}) is another kind of 
generalization of the equilibrium detailed balance condition  
to the nonequilibrium steady state, like Eq. 
(\ref{DetaiBalan1}).

   We introduce the distribution function $P_{r}(R,t)$ of 
the dimensionless energy loss $R$ by friction 
in the time-interval $[t_{0},t]$ as   
\begin{eqnarray}
   P_{r}(R,t) = \pathaveA{\delta\!\left(R - 
   \beta\mathcal{R}_{t}^{(v)}(y_{t},y_{0})\right) }
\label{FrictDistr1}\end{eqnarray} 
   Like for the work distribution function, we represent 
the distribution function of energy loss by friction in the form 
\begin{eqnarray}
   P_{r}(R,t) 
   = \frac{1}{2\pi}\int_{-\infty}^{+\infty}d\lambda \;
   e^{i\lambda R} \mathcal{E}_{r}(i\lambda,t)
\label{DistrFrict2}\end{eqnarray}
where the Fourier transformation 
$\mathcal{E}_{r}(i\lambda,t)$ is given by 
\begin{eqnarray}
   \mathcal{E}_{r}(\lambda,t) &\equiv& 
   \pathaveA{e^{-\lambda
   \beta\mathcal{R}_{t}^{(v)}(y_{t},y_{0})}} 
   \label{EFunctFrict1} \\
   &=& 
   \int dy_{t} \! \int dy_{0} \;   
      \transP{F}{y_{t}}{t}{y_{0}}{t_{0}} 
      e^{-\lambda \beta\mathcal{R}_{t}^{(v)}(y_{t},y_{0}) } 
      \nonumber \\
   &&\spaEq \times  
      f(y_{0},t_{0}) 
\label{EFunctFrict2}\end{eqnarray}
%. 
with Eqs. (\ref{TransProba1}) and (\ref{PathAvera1}).
   Here, the $v$-dependence of the function 
$\mathcal{E}_{r}(\lambda,t)$ for friction, 
as well as a similar $\mathcal{E}$-function for heat introduced later, 
has been suppressed. 
   It follows from Eqs. (\ref{DetaiBalan4}), (\ref{EFunctFrict2}) 
and $\mathcal{R}_{t}^{(v)}(y_{0},y_{t}) 
= - \mathcal{R}_{t}^{(v)}(y_{t},y_{0})$ that 
\begin{eqnarray}
   \mathcal{E}_{r}(1-\lambda,t) = \mathcal{E}_{r}(\lambda,t)  
\label{FluctTheorFrict1}\end{eqnarray}
\textit{if} $f(y_{0},t_{0})=f_{eq}(y_{0})$. 
   Or equivalently, for the distribution function $P_{r}(R,t)$ of 
the dimensionless energy loss by friction, 
using Eqs. (\ref{DistrFrict2}) 
and (\ref{FluctTheorFrict1}) we obtain  
\begin{eqnarray}
   \frac{P_{r}(R,t)}{P_{r}(-R,t)} = \exp(R) 
\label{FluctTheorFrict2}\end{eqnarray}
for the equilibrium initial condition. 
   Equation (\ref{FluctTheorFrict2}) is  
the transient fluctuation theorem for friction and is   
satisfied for any time $t$.

   If one is interested in the derivation of a fluctuation 
theorem for more general initial states 
than the equilibrium initial state, 
we can proceed as follows. 
%   Now, we investigate what happens if the system at the initial time 
%is not in the equilibrium state. 
%   Inserting Eqs. (\ref{TransProba2}) and (\ref{FrictLoss}) 
%into Eq. (\ref{EFunctFrict2}), and then using Eq. (\ref{DistrFrict2}) 
%we obtain 
   Using Eqs. (\ref{TransProba1}), 
(\ref{PathAvera1}) and (\ref{FrictDistr1}) we obtain 
\begin{widetext}
\begin{eqnarray}
   P_{r}(R,t)
   &=& \int dy_{t}\! \int dy_{0} \; f(y_{0},t_{0})
   \; \delta\!\left(R 
   - \beta \mathcal{R}_{t}^{(v)}(y_{t},y_{0}) \right) 
%c      \nonumber \\
%c   &&\spaEq\times   
   \transP{F}{y_{t}}{t}{y_{0}}{t_{0}} 
      \nonumber \\
%   &=& \frac{D}{|v|} \int dy_{0} \; 
%      \transP{F}{y_{0}-\frac{DR}{v}}{t}{y_{0}}{t_{0}} 
%      f(y_{0},t_{0}) 
%      \nonumber \\
   &=&  \frac{1}{\sqrt{4\pi\alpha\beta v^{2}\mathcal{T}_{t}}} 
      \int dy_{0}\;f(y_{0},t_{0}) 
%      \nonumber \\
%   && \spaEq\times  
      \exp \left\{ - \frac{ \left[R-\alpha\beta v(y_{0} +v\tau)
      \left(1-b_{t}\right) \right]^{2}}
      {4\alpha\beta v^{2}\mathcal{T}_{t}} 
      \right\}  
\label{DistrFrict3}\end{eqnarray}
\end{widetext}
for \emph{any} initial distribution $f(y_{0},t_{0})$,  
where we used Eqs.  (\ref{TransProba2}), (\ref{FrictLoss}) and 
$ \delta (R - \beta \mathcal{R}_{t}^{(v)}(y_{t},y_{0}) ) = 
   \delta (y_{t} - y_{0}+ R/(\alpha \beta v) ) 
   /(\alpha \beta |v|)$. 
%   [A derivation of Eq. (\ref{DistrFrict3}) is in 
%Appendix \ref{CalculationDistributionFriction}.] 

   To get more concrete results, 
in the remaining part of this section 
we concentrate on the initial 
distribution 
\begin{eqnarray}
   f(y_{0},t_{0})=f_{eq}(y_{0}+v\tau\phi),
\label{InitiCondi1}\end{eqnarray}
for a constant parameter $\phi$, giving the 
equilibrium initial distribution for $\phi=0$ 
and the non-equilibrium steady state initial distribution 
for $\phi=1$. 
   Inserting Eq. (\ref{InitiCondi1}) into 
Eq. (\ref{DistrFrict3}) the distribution $P_{r}(R,t)$ is given by 
\begin{eqnarray}
   &&\hspace{-0.5cm}
   P_{r}(R,t) 
   \nonumber \\
   &&
   = \frac{1}{\sqrt{4\pi\alpha\beta v^{2}\tau
      \left(1-b_{t}\right)}}
      \nonumber \\
   && \spaEq\times
      \exp \left\{  - \frac{
      \left[R - \alpha\beta v^{2}\tau(1-\phi)
      \left(1-b_{t}\right) 
      \right]^{2}}{4 \alpha\beta v^{2} \tau
      \left(1-b_{t}\right) } 
      \right\} \;\;\;
\label{DistrFrict5}\end{eqnarray}
using Eq. (\ref{EquilDistr1}). 
  It follows from Eq. (\ref{DistrFrict5}) that   
\begin{eqnarray}
 %  \lim_{t\rightarrow+\infty}
   \frac{P_{r}(R,t)}{P_{r}(-R,t)} 
   = \exp[(1-\phi)R],  
\label{FluctTheorFrict3}\end{eqnarray}
which does not have the 
form of a fluctuation theorem for $\phi\neq 0$.  
%ap    [we give  derivation of Eq. (\ref{FluctTheorFrict3}) in 
%ap Appendix \ref{CalculationDistributionFriction}.] 
   In other words, the distribution function of energy loss  
by friction satisfies the transient fluctuation theorem 
for $\phi=0$, 
but not the steady state fluctuation theorem using 
the steady state initial condition (\ref{InitiCondi1}) 
for $\phi=1$. 
   Actually, for the initial condition 
of a nonequilibrium steady state, i.e. if $\phi=1$, 
its distribution $P_{r}(R,t)$ 
is Gaussian from Eq. (\ref{DistrFrict5}) 
with its peak at $R=0$, therefore 
$P_{r}(-R,t)=P_{r}(R,t)$ then at any time.

%%%%%%%%%%%%%%%%%%%%%%%%%%%%%%%%%%%%%%%%%%%%%%%%%%%%%%%%%%%%%%%%%%%%%%
\section{Extended Fluctuation Theorem for Heat}
\label{FluctuationTheoremHeat}

   As the next topic of this paper, we consider  
the distribution function of heat, which was  
defined in Sec. \ref{HeatEntropyBalance}, 
but now we calculate it by carrying out a functional integral.   
and also discuss a new simple derivation of  
its fluctuation theorem briefly in the long time limit.

%---------------------------------------------------------------------
%\subsection{Functional integral calculation of the heat distribution function}

   The distribution function of the dimensionless heat $Q$  
corresponding to $\beta\mathcal{Q}^{(v)}_{t}(\{y_{s}\})$ using Eq. 
(\ref{Heat1}) is given by   
\begin{eqnarray}
   P_{q}(Q,t) = \pathaveA{\delta\!\left(Q - 
   \beta\mathcal{Q}^{(v)}_{t}(\{y_{s}\})\right)} .
\label{HeatDistr1}\end{eqnarray}
   The heat distribution function $P_{q}(Q,t)$ can be calculated 
like in the distribution function of work or energy loss by friction, 
namely by representing it as 
\begin{eqnarray}
   P_{q}(Q,t) 
   = \frac{1}{2\pi}\int_{-\infty}^{+\infty}d\lambda \;
   e^{i\lambda Q} \mathcal{E}_{q}(i\lambda,t)
\label{HeatDistr2}\end{eqnarray}
where $\mathcal{E}_{q}(\lambda,t)$ is given by 
\begin{eqnarray}
   \mathcal{E}_{q}(\lambda,t) &\equiv& 
      \pathaveA{e^{-\lambda \beta\mathcal{Q}^{(v)}_{t}(\{y_{s}\}) } }
      \label{EFunctHeat1} \\
   &=& \int dy_{t}\! \int dy_{0} \; e^{\lambda \beta U(y_{t}) }
      \mathcal{F}(y_{t},y_{0};\lambda)  e^{-\lambda \beta U(y_{0}) }
      \nonumber \\
   &&\spaEq \times
      f(y_{0},t_{0}) 
\label{EFunctHeat2}\end{eqnarray}
where we used Eqs. (\ref{EnergyBalan1}), 
(\ref{EnergDiffe1}), (\ref{PathAvera1}) 
and (\ref{EFunctWorkTrnas1}) to derive Eq. 
(\ref{EFunctHeat2}) from Eq. (\ref{EFunctHeat1}). 
%   Here, the $v$-dependence of the function 
%$\mathcal{E}_{q}(\lambda,t)$ has been suppressed 
%here and in the rest of the paper, like $\mathcal{E}_{r}(\lambda,t)$.  
   It may be meaningful to notice that from Eqs. 
(\ref{EFunctWork2}) and (\ref{EFunctHeat2}) 
the function $\mathcal{E}_{q}(\lambda,t)$ for heat 
is different from the function $\mathcal{E}_{w}^{(v)}(\lambda,t)$ 
for work by the factor $\exp\{\lambda \beta [U(y_{t})-U(y_{0})] \}$ 
only. 
% using the function $\mathcal{F}(y_{t},y_{0};\lambda)$ 
% given by Eq. (\ref{EFunctWorkTrnas1}). 
%    Using Eq. (\ref{EFunctWorkTrnas2}) with 
% Eqs. (\ref{EularLagra2Solve1}) and (\ref{PathInteg1}), 
   Inserting Eq. (\ref{FunctCalF1}) into  
Eq. (\ref{EFunctHeat2}) one obtains
\begin{widetext}
\begin{eqnarray}
   \mathcal{E}_{q}(\lambda,t) 
   &=& \frac{1}{\sqrt{1 - \lambda \left(1-b_{t}^{2}\right)}}
      \exp\left[ - \lambda (1-\lambda) \alpha \beta v^{2} 
      \left( t-t_{0} - 2\tau \frac{1-b_{t}}{1+b_{t}} \right)  \right]
      \nonumber \\
   &&\spaEq \times \int dy_{0} \;  f(y_{0},t_{0}) 
      \exp\left[ - \frac{\beta\kappa}{2}
      \frac{\lambda(1 - \lambda )\left(1-b_{t}^{2}\right) }
      {1 - \lambda \left(1-b_{t}^{2}\right)} 
      \left(y_{0} - v\tau \frac{1-b_{t}}{1+b_{t}}  \right)^{2}
      \right] 
\label{EFunctHeat3}\end{eqnarray}
\end{widetext}
for any initial distribution $f(y_{0},t_{0})$. 
%ap   [We give a derivation of Eq. (\ref{EFunctHeat3}) in 
%ap Appendix \ref{CalculationDistributionHeat}.]

   The calculation of the heat distribution function $P_{q}(Q,t)$ 
from its Fourier transformation like 
$\mathcal{E}_{q}(i\lambda,t)$ has already done in Ref. \cite{ZC04} 
in detail, for the initial condition of the equilibrium 
and nonequilibrium steady state, 
and led to the extended fluctuation theorem for heat 
\cite{ZC03a,ZC04}.    
   We do not repeat their calculations and argument in this paper.  
%and simply suggest readers to read Ref. \cite{ZC04} for the 
%calculation of the heat distribution function $P_{q}(Q,t)$ 
%from $\mathcal{E}_{q}(i\lambda,t)$ and for arguments about  
%the extended fluctuation theorem for heats using 
%the distribution function $P_{q}(Q,t)$. 
   Instead, in the remaining of this section we discuss 
the heat distribution function and its fluctuation theorem 
by a less rigorous but much simpler argument than 
in Ref. \cite{ZC04}.  
   This discussion is restricted to the case 
of the long time limit, 
in which time-correlations of some quantities may be neglected.  
   This allows to simplify considerably 
the derivation of the relevant distribution functions. 
   The heat fluctuation theorem is also discussed in Refs. 
\cite{BGG05,G06,BJM06}.

%---------------------------------------------------------------------

%   In the remaining part of this section, we consider 
%the case of the initial condition (\ref{InitiCondi1}) again. 
%   First, we consider the distribution function 
%$P_{e}(E)$ of the dimensionless potential energy $E$ 
%for the initial case 
%of $f_{eq}(y+v\tau\phi)$, given by  
%
%\begin{eqnarray}
%   P_{e}^{(\phi)}(E) &\equiv&\int dy\; 
%      f_{eq}(y+v\tau\phi) \delta (\beta U(y) - E) 
%      \nonumber \\
%   &=& \frac{\theta(E)}{2\sqrt{\pi E}} \left[e^{-\left(\sqrt{E}
%      +v\tau\phi\sqrt{\frac{\beta\kappa}{2}}\right)^{2}}
%%c      \right.\nonumber \\
%%c   &&\spaEq\spaEq\left.
%      +e^{-\left(\sqrt{E}
%      -v\tau\phi\sqrt{\frac{\beta\kappa}{2}}\right)^{2}} \right]
%\label{EnergDistri1}\end{eqnarray}
%
%where we used $U(y)=\kappa y^{2}/2$ so that 
%$\delta (\beta U(y) - E) = \theta(E) 
%[\delta (y+\sqrt{2E/(\beta\kappa)}) 
%+ \delta (y-\sqrt{2E/(\beta\kappa)})]/\sqrt{2\beta\kappa E}$, 
%and  $\theta (x)$ is the Heaviside function 
%taking the value $1$ for $x>0$ and $0$ for $x \leq 0$. 
%
%\begin{eqnarray}
%   \theta (x) \equiv 
%   \left\{\begin{array}{ll}
%      1 & \mbox{in}\; x > 0 \\
%      0 & \mbox{in}\; x \leq 0 .
%   \end{array}\right.
%\label{Heaviside}\end{eqnarray} 
%
%   Now, we consider the case of 
%%$|2E/(\beta\kappa v^{2}\tau^{2})| = |2DE/(v^{2}\tau)|>\!>1$
%$|E|>\!> \beta\kappa v^{2}\tau^{2}/2$
%[or $|E|>\!>v^{2}\tau/(2D)$] 
%in Eq. (\ref{EnergDistri1}), 
%and approximate the distribution function $P_{e}(E)$ as 
%
   We start our argument by assuming that the particle energy 
is canonical-like distributed 
due to the presence of the fluid surrounding a Brownian particle 
\cite{ZC03a}, 
so that the distribution $P_{e}(E)$ 
of the dimensionless energy $E$, 
i.e. the (potential) energy times the inverse temperature $\beta$,  
is given by 
\begin{eqnarray}
   P_{e} (E) \approx \theta(E) \exp(-E) ,
\label{EnergDistri2}\end{eqnarray}
where $\theta (x)$ is the Heaviside function 
taking the value $1$ for $x>0$ and $0$ for $x \leq 0$, 
and $\theta(E)$ in Eq. (\ref{EnergDistri2}) guarantees 
that the energy $E$ is positive. 
%neglecting the power dependence $E^{-1/2}$ 
%in Eq. (\ref{EnergDistri1}), which is less strong 
%than the exponential dependence $\exp(-E)$.  
  [Note that on the right-hand side of Eq. (\ref{EnergDistri2}) 
the normalization condition 
$\int dE P_{e}(E)=1$ is still satisfied.]  
%and $\phi$-dependence of $P_{e}^{(\phi)}(E)$  
%in Eq. (\ref{EnergDistri2}) disappears  in this approximation.] 
   Now, we consider the distribution function 
$P_{\Delta e}(\Delta E,t)$ of the dimensionless energy difference 
$\Delta E (=E_{t} - E_{0})$ at the initial time $t_{0}$ and 
the final time $t$, which is given by 
\begin{eqnarray}
   P_{\Delta e}(\Delta E,t) 
   &\stackrel{t\rightarrow+\infty}{\sim}& 
   \int d E_{0}\int d E_{t}\; 
   P_{e}(E_{0}) 
   P_{e}(E_{t}) 
   \nonumber \\ 
   &&\spaEq \times
   \delta(E_{t} - E_{0} - \Delta E) ,  
\label{EnergDiffeDistr1}\end{eqnarray}
namely $\int d E_{0}\; P_{e}(E_{0}) 
P_{e}(E_{0}+\Delta E)$, 
in the long-time limit. 
%under the initial condition  (\ref{InitiCondi1}). 
   Here, we have assumed that the energy $E_{0}$ at the 
initial time $t_{0}$ and the energy 
$E_{t}$ %\equiv E_{0}+\Delta E$  
at the final time $t$ are uncorrelated in the long time limit 
$t\rightarrow+\infty$, so that 
the distribution function of the energies $E_{0}$ and 
$E_{t}$ is given by 
a multiplication of $P_{e}(E_{0})$ 
(the initial energy probability distribution) and  
$P_{e}(E_{t})$
(the final energy probability distribution).  
%noting that the asymptotic state is $\phi =1$). 
   The distribution function 
$P_{\Delta e}(\Delta E,t)$ is given by the integral of 
such a distribution function of the energies $E_{0}$ and $E_{t}$ 
%the simple 
%multiplication $P_{e}^{(\phi)}(E_{0}) P_{e}^{(1)}(E_{t})$ 
%of the initial distribution function $P_{e}^{(\phi)}(E_{0})$ 
%of energy and the final distribution function $P_{e}^{(1)}(E_{t})$ 
%of energy (noting that the asymptotic state is $\phi =1$) 
over all possible values of $E_{0}$ and $E_{t}$ under the 
constraint $\Delta E = E_{t}-E_{0}$, 
therefore by Eq. (\ref{EnergDiffeDistr1}). 
%   It should be noted that the asymptotic limit 
%$t\rightarrow+\infty$ 
%in Eq. (\ref{EnergDiffeDistr1}) physically implies  
%the condition $t-t_{0} >\!> \tau$.    
%and this condition and the assumption   
%$|E|>\!> \beta\kappa v^{2}\tau^{2}/2$
%to derive Eq. (\ref{EnergDistri2})  
%can both be satisfied by a small $\tau$.
   Inserting Eq. (\ref{EnergDistri2}) 
into Eq. (\ref{EnergDiffeDistr1}) we obtain 
\begin{eqnarray}
   P_{\Delta e}(\Delta E,t) 
   \stackrel{t\rightarrow+\infty}{\sim} 
   \frac{1}{2}\exp(-|\Delta E|) ,
\label{EnergDiffeDistr2}\end{eqnarray}
meaning that the distribution function $P_{\Delta e}(\Delta E,t)$ 
of the energy difference $\Delta E$ decays exponentially. 
   [Note again that the right-hand side of (\ref{EnergDiffeDistr2}) 
satisfies the normalization condition 
$\int d\Delta E P_{\Delta e}(\Delta E,t) = 1$.]
   The argument leading to Eq. (\ref{EnergDiffeDistr2}) 
is also in Ref. \cite{ZC03a}. 

   On the other hand, we have already calculated the distribution 
function $P_{w}(W,t)$ of work $W$ in 
Sec. \ref{FunctionalCalculationWorkDistribution} and 
from Eq. (\ref{DistrWorkLongTime1}) we derive
\begin{eqnarray}
   P_{w}(W,t) \stackrel{t\rightarrow+\infty}{\sim}
   \frac{1}{\sqrt{4\pi\overline{W}_{t}}} 
   \exp\left[-\frac{\left(W-  \overline{W}_{t} \right)^{2}}
   {4  \overline{W}_{t}} \right] 
\label{WorkDistr4}\end{eqnarray} 
for any initial distribution. 
%including the case (\ref{InitiCondi1}). 
   Here, $\overline{W}_{t}$ is the average of the (dimensionless) 
work $W$ for the distribution $P_{w}(W,t)$ and given 
by $\overline{W}_{t}\stackrel{t\rightarrow+\infty}{\sim}
\alpha\beta v^{2} t$ from Eq. (\ref{DistrWorkLongTime1}) 
in the long time limit. 

   By Eq. (\ref{EnergyBalan1}), 
%in Sec. \ref{HeatEntropyBalance}, 
the heat $Q$ is given by $Q=W-\Delta E$ using the work $W$ 
and energy difference $\Delta E$, and its distribution 
function $P_{q}(Q,t)$ should be represented as  
\begin{eqnarray}
   P_{q}(Q,t) & 
   \stackrel{t\rightarrow+\infty}{\sim}& 
   \int d W \int d \Delta E\; P_{w}(W,t) P_{\Delta e}(\Delta E,t)
   \nonumber \\ 
   &&\spaEq \times \delta (W-\Delta E - Q), 
\label{HeatDistr3}\end{eqnarray}
namely $\int d W \; P_{w}(W,t) P_{\Delta e}(W-Q,t)$, 
in the long time limit. 
   Here, we used a similar argument 
as in Eq. (\ref{EnergDiffeDistr1}) 
in order to justify Eq. (\ref{HeatDistr3}), 
namely, Eq. (\ref{HeatDistr3}) is the integral 
of the multiplication 
of the work distribution 
$P_{w}(W,t)$ and the energy-difference distribution function 
$P_{\Delta e}(\Delta E,t)$ over all possible values of $W$ 
and $\Delta E$ 
under the restriction $Q=W-\Delta E$ for a given $\Delta E$. 
    Non-correlation of the work and the energy difference 
in the long time limit, which is assumed in Eq. (\ref{HeatDistr3}), 
may be justified by the fact that the work depends on the 
particle position over the entire time interval $[t_{0},t]$ 
by Eqs. (\ref{WorkRate1}) and (\ref{Work1}) 
(in which the effects at the times $t_{0}$ and $t$ are 
negligible in the long time limit) while the energy difference 
depends exclusively 
on the particle positions at the times $t_{0}$ and $t$ only.  
   Inserting Eq. (\ref{EnergDiffeDistr2}) and (\ref{WorkDistr4}) 
into Eq. (\ref{HeatDistr3}) we obtain 
\begin{eqnarray}
   P_{q}(Q,t) 
   &\stackrel{t\rightarrow+\infty}{\sim}& \frac{1}{4} \left[
      e^{-Q+2\overline{W}_{t}}
      \mbox{erfc}\left(-\frac{Q- 3 \overline{W}_{t}}
      {2\sqrt{\overline{W}_{t}}}\right)
      \right.\nonumber \\
   &&\spaEq \left.
      +  e^{Q}\mbox{erfc}\left(\frac{Q + \overline{W}_{t}}
      {2\sqrt{\overline{W}_{t}}} \right) \right] 
\label{HeatDistr4}\end{eqnarray}
with the complimentary error function $\mbox{erfc}(x)$ 
defined by 
%
%\begin{eqnarray}
$   \mbox{erfc}(x) \equiv (2/\sqrt{\pi}) 
   \int_{x}^{+\infty} dz\; \exp(-z^{2}) $, 
%\label{ErrorFunct1}\end{eqnarray} 
satisfying the inequality $0<\mbox{erfc}(x)<2$.
   One may notice that the average work  $\overline{W}_{t}$ 
is equal to the average heat in the case of the nonequilibrium 
steady state initial condition $f(y_{0},t_{0}) 
=f_{ss}(y_{0})$, because of the  
energy conservation law (\ref{EnergyBalan1}) and 
the fact that average of the internal energy 
difference (\ref{EnergDiffe1}) is zero in this case. 
   Equation (\ref{HeatDistr4}) gives the asymptotic form 
of the heat distribution function. 
   The exponential factors $\exp(\pm Q)$ in Eq. (\ref{HeatDistr4}) 
dominate the tails of the heat distribution function $P_{q}(Q,t)$ 
\cite{ZC04}.

%
%---------------------------------------------------------------------
\begin{figure}[!t]
\vspfigA
\begin{center}
\includegraphics[width=\widthfig]{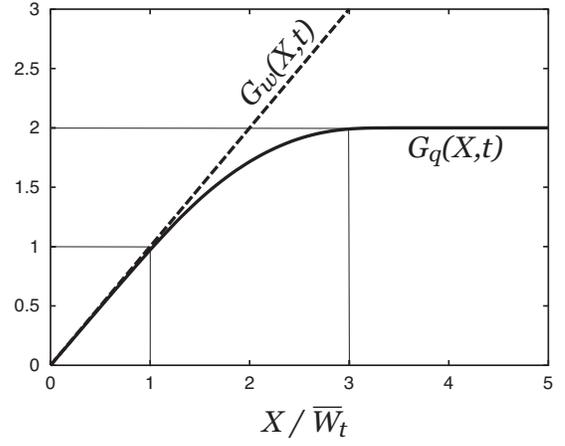}
\vspfigB
\caption{ Comparison of the work fluctuation theorem and 
   the heat fluctuation theorem  
   by plotting the function $G_{w}(X,t)$ 
   for the work distribution and 
   the function $G_{q}(X,t)$ 
   for the heat distribution as functions of 
   a scaled $X/\overline{W}_{t}$ 
   %($X=W$ or $Q$) 
   in the long time limit $t\rightarrow +\infty$. 
      Here, we used the asymptotic forms  (\ref{WorkDistr4}) 
   and (\ref{HeatDistr4}) 
   of the work distribution function and 
   the heat distribution function, respectively, 
   in the case of $\overline{W}_{t}=70$.  
      In the small $X$ region 
   ($0\leq X/\overline{W}_{t} \leq 1$), 
   the values of the 
   two functions $G_{w}(X,t)$ and $G_{q}(X,t)$ 
   appear to be consistent with $G_{w}(X,t) 
   =G_{q}(X,t)$, 
   while $G_{w}(X,t)$ is $X/ \overline{W}_{t}$ 
   and $G_{q}(X,t)$ is 2 in the large $X$ region 
   ($3\leq X/\overline{W}_{t} \leq +\infty$) 
   in the long time limit. 
 }
\label{figB1heatFT}
\end{center}
\vspfigC
\end{figure}  
%--------------------------------------------------------------------- 
%
%   Now, we introduce the normalized variables 
%$p_{w}$ and $p_{q}$ of $W$ and $Q$ by $\overline{W}_{t}$ as  
%
%\begin{eqnarray}
%$   p_{w} \equiv W/\overline{W}_{t}$  and  
%   $p_{q} \equiv Q/\overline{W}_{t}$ ,
%\label{}
%\end{eqnarray}
%

   Now, we introduce the fluctuation functions 
$G_{w}(W,t)$ and $G_{q}(Q,t)$ 
defined by 
\begin{eqnarray}
   G_{w}(W,t) &\equiv& 
      \frac{1}{ \overline{W}_{t}} \ln\frac{P_{w}(W,t)}
      {P_{w}(-W,t)} , 
      \label{WorkFluctFunct1}\\
   G_{q}(Q,t) &\equiv& 
      \frac{1}{ \overline{W}_{t}}  \ln\frac{P_{q}(Q,t)}
      {P_{q}(-Q,t)} 
      \label{HeatFluctFunct1}
\end{eqnarray}
for the work distribution function 
$P_{w}(W,t)$ and the heat distribution function $P_{q}(Q,t)$. 
   By Eq. (\ref{WorkDistr4}) the function 
$G_{w}(W,t)$ is given simply by 
$G_{w}(W,t)\stackrel{t\rightarrow+\infty}{\sim}W/ \overline{W}_{t}$ 
in the long time limit, characterizing 
the work fluctuation theorem in a proper way to compare it 
with the heat fluctuation theorem characterized by 
the function $G_{q}(Q,t)$. 
   In Fig. \ref{figB1heatFT} the functions 
$G_{w}(W,t)$ (broken line) and $G_{q}(Q,t)$ (solid line) 
are plotted as functions of $X/\overline{W}_{t}$ ($X=W$ or $Q$)
using Eqs. (\ref{WorkDistr4}) and (\ref{HeatDistr4}) 
in the case of $\overline{W}_{t}=70$. 
   In this figure we plotted only in the positive region 
of $W$ and $Q$, because their values in 
the negative region is simply given by 
$G_{w}(-W,t) = -G_{w}(W,t)$ and 
$G_{q}(-Q,t) = - G_{q}(Q,t)$. 
   It is clear from Fig. \ref{figB1heatFT} that 
the values of the functions $G_{w}(X,t)$ and $G_{q}(X,t)$ 
will coincide with each other for small $W$ and $Q$ 
for $\overline{W}_{t}\rightarrow +\infty$, 
i.e. $t\rightarrow +\infty$, meaning that 
the heat fluctuation theorem coincides 
with the work fluctuation theorem in this region. 
   The difference between the heat and work fluctuation theorems 
appears in the large values of the argument, 
where the function $G_{w}(W,t)$ 
remains $W/\overline{W}_{t}$, 
while the function $G_{q}(Q,t)$ takes the constant 
value $2$ for $Q/\overline{W}_{t} > 3$ in the long time limit. 
   For further details, we refer to Ref. \cite{ZC04}.

%%%%%%%%%%%%%%%%%%%%%%%%%%%%%%%%%%%%%%%%%%%%%%%%%%%%%%%%%%%%%%%%%%%%%%
\section{Inertial Effects}
\label{InertiaEffects}

   So far, we have concentrated our discussions 
to the over-damped case and have neglected inertial effects. 
   A generalization of our discussions to the ones including 
the inertia is almost straightforward.  
   One of the features caused by introducing inertia is 
a kinetic term in the equilibrium and nonequilibrium 
steady state distribution functions. 
   This kinetic term depends on the frame one uses, namely 
the comoving frame or the laboratory frame, respectively. 
   The inertial force, like d'Alembert's force, 
also appears as an inertial effect. 
   In this section we discuss briefly these effects 
beyond the over-damped case. 

   The Langevin equation including inertia is expressed as 
Eq. (\ref{LangeEq3}) in the laboratory frame. 
%   [In this section, 
%for the inertia case we use the same notations 
%for the particle position like $x_{t}$ and $y_{t}$ 
%as for the over-damped case.]
   Like in the over-damped case, we can convert Eq. 
(\ref{LangeEq3}) for the laboratory frame to 
\begin{eqnarray}
   m\frac{d^{2} y_{t}}{dt^{2}} 
   = -\alpha \frac{d y_{t}}{dt}
      - \kappa  \left(y_{t} +v\tau\right)+ \zeta_{t} 
\label{LangeEq4}\end{eqnarray}
for the comoving frame by Eq. (\ref{ComovPosit1}). 
   Equation (\ref{LangeEq4}) reduces to  
Eq. (\ref{LangeEq2}) for the over-damped case 
when $m d^{2} y_{t}/dt^{2}=0$. 

   We introduce a canonical-like distribution function as 
\begin{eqnarray}
   f_{eq}^{(\vartheta)}(\dot{y},y) 
   \equiv \Xi^{(\vartheta)}{}^{-1}\exp\left[-\beta 
   \mathcal{H}\left(\dot{y}+\vartheta v,y\right)\right]
\label{EquilDistr2}\end{eqnarray}
where $\dot{y}$ is the time-derivative of $y$ and 
$\mathcal{H}\left(\dot{y},y\right)$ is defined by 
\begin{eqnarray}
%$
   \mathcal{H}\left(\dot{y},y\right) 
   \equiv m\dot{y}^{2}/2 + \kappa y^{2}/2, 
\label{HamilFunct1}\end{eqnarray}
and $\Xi^{(\vartheta)}$ is the normalization constant for 
the distribution function $f_{eq}^{(\vartheta)}(\dot{y},y)$. 
   It is important to note that the particle velocity 
depends on the frame, and is given by $\dot{y}$ 
for the comoving frame and by $\dot{x}(=\dot{y}+v)$ for 
the laboratory frame. 
   For that reason, the canonical distribution function 
$f_{eq}^{(\vartheta)}(\dot{y},y)$ including the kinetic 
energy depends on the frame, 
so that $f_{eq}^{(0)}(\dot{y},y)$ for $\vartheta=0$ 
refers to the comoving frame 
and $f_{eq}^{(1)}(\dot{y},y)$ for $\vartheta=1$ 
refers to the laboratory frame. 

   By a way similar to the over-damped case, the functional 
probability density for path $\{y_{s}\}_{s\in[t_{0},t]}$ 
is given by $\exp[\int_{t_{0}}^{t} ds\; 
L^{(v)}(\ddot{y}_{s},\dot{y}_{s},y_{s})]$ with 
the Lagrangian function 
\begin{eqnarray}
    L^{(v)}\!\left(\ddot{y}_{s},\dot{y}_{s},y_{s}\right) 
    \equiv -\frac{1}{4D}\left(
    \dot{y}_{s} 
    +\frac{1}{\tau} y_{s} +v 
    +\frac{m}{\alpha}\ddot{y}_{s}
    \right)^{2} 
\label{Lagra3}\end{eqnarray}
using $\ddot{y}_{s}\equiv d^{2}y_{s}/ds^{2}$. 
   The Lagrangian function (\ref{Lagra3}) becomes  
the Lagrangian function (\ref{Lagra1}) 
in the over-damped case, where  $m\ddot{y}_{s}=0$. 
   Using Eqs. (\ref{EquilDistr2}) and (\ref{Lagra3}) 
we obtain 
\begin{eqnarray}
   &&
   e^{-\beta \int_{t_{0}}^{t}ds\; 
      \Lambda_{\pm}(\ddot{y}_{s},\dot{y}_{s},y_{s};\vartheta) v}
   e^{\int_{t_{0}}^{t} ds\; 
      L^{(v)}\!\left(\ddot{y}_{s},\dot{y}_{s},y_{s}\right)} 
      f_{eq}^{(\vartheta)}\!\left(\dot{y}_{0},y_{0}\right)
      \nonumber \\
   &&\spaEq 
   = f_{eq}^{(\vartheta)}\!\left(\dot{y}_{t},y_{t}\right) 
      e^{\int_{t_{0}}^{t} ds\; 
      L^{(\pm v)}\left(\ddot{y}_{s},-\dot{y}_{s},y_{s}\right)} 
%c     \nonumber \\
%c   &&\spaEq\spaEq \times , 
\label{DetaiBalan7}\end{eqnarray}
where $\dot{y}_{0}=\dot{y}_{t_{0}}$, 
and $\Lambda_{\pm}(\ddot{y}_{s},\dot{y}_{s},y_{s};\vartheta)$
is a modified ``force'' defined by 
\begin{eqnarray}
   \Lambda_{\pm}(\ddot{y}_{s},\dot{y}_{s},y_{s};\vartheta)
      &\equiv& 
      %\equiv
      - \kappa y_{s} \frac{1\mp 1}{2} 
      - \alpha  \dot{y}_{s}\frac{1\pm 1}{2}
      \nonumber \\
   && \spaEq
      - m\ddot{y}_{s} \left(\frac{1\mp 1}{2}-\vartheta\right).
\label{GenerForce1}\end{eqnarray}
   Equation (\ref{DetaiBalan7}) may be regarded as a nonequilibrium  
detailed balance relation for the case 
of a potential force, friction and inertia. 
   [See Appendix \ref{NonequilibriumDetailedBalanceInertia} 
for a derivation of Eq. (\ref{DetaiBalan7}).]
%   Here, $\vartheta=0$ ($\vartheta=1$) is for the comoving frame 
%(the laboratory frame). 
   Moreover, the signs $\pm$ 
in Eq. (\ref{DetaiBalan7}) correspond to the case of work ($-1$),  
discussed in \ref{FluctuationTheoremWork} 
and that of energy loss by friction ($+1$), respectively, 
discussed in Sec. \ref{FluctuationTheoremFriction}, 
and are due to the $\pm v$ signs of the Lagrangian 
$L^{(\pm v)} (\ddot{y}_{s},-\dot{y}_{s},y_{s} )$ 
on the right-hand side of Eq. (\ref{DetaiBalan7}).
   It should be noted that the first, second and third terms 
on the right-hand side of Eq. (\ref{GenerForce1}) are regarded as 
the harmonic force, 
the friction force, and the inertial 
(d'Alembert-like) force, respectively.

%
%---------------------------------------------------------------------
%\begin{widetext}
%\begin{center}
\begin{table*}
\vspfigA
\renewcommand{\arraystretch}{2.1}
\caption{
      Four kinds of fluctuation theorems corresponding 
   to four forces in the case including inertial effects.
      Here, $\vartheta=0$ ($\vartheta=1$) is for the case of  
   the comoving frame (the laboratory frame), 
   and $\pm v$ is the velocity appearing in the 
   Lagrangian function 
   $L^{(\pm v)}(\ddot{y}_{s},-\dot{y}_{s},y_{s})$ 
   on the right-hand side 
   of the nonequilibrium detailed balance relation 
   (\ref{DetaiBalan7}).}
\label{4FluctTheor}
\vspfigB
\vspace{0.5cm}
\begin{tabular}{c|c|l|c}
   \hspace{0.8cm} Frame ($\vartheta$)\hspace{0.8cm} &
   $\hspace{0.5cm}\pm v\hspace{0.5cm}$ & 
   \hspace{2cm}Force $\Lambda$  \hspace{2cm} & 
   \hspace{0.5cm} Fluctuation Theorem   \hspace{0.5cm} \\
   \hline
%
%&\\
   Comoving ($0$) &
   $-v$ & \hspace{0.5cm} $\Lambda_{-}(\ddot{y},\dot{y},y;0) 
      =-\kappa y - m\ddot{y}$ & 
   $\textstyle \frac{\mathcal{P}_{-}^{(0)}(\mathcal{W},t)}
   {\mathcal{P}_{-}^{(0)}(-\mathcal{W},t)} 
   = \exp(\mathcal{W})$  \\
%
%&\\
   Comoving ($0$) &
   $+v$ & \hspace{0.5cm} $\Lambda_{+}(\ddot{y},\dot{y},y;0) 
      =  -  \alpha  \dot{y}  $ & 
   $\textstyle \frac{\mathcal{P}_{+}^{(0)}(\mathcal{W},t)}
   {\mathcal{P}_{+}^{(0)}(-\mathcal{W},t)} 
   = \exp(\mathcal{W})$  \\
%
%&\\
   Laboratory ($1$) &
   $-v$ & \hspace{0.5cm} $\Lambda_{-}(\ddot{y},\dot{y},y;1) 
      = - \kappa y$  & 
   $\textstyle \frac{\mathcal{P}_{-}^{(1)}(\mathcal{W},t)}
   {\mathcal{P}_{-}^{(1)}(-\mathcal{W},t)} 
   = \exp(\mathcal{W})$  \\
%
%&\\
   Laboratory ($1$) &
   $+v$ & \hspace{0.5cm} $\Lambda_{+}(\ddot{y},\dot{y},y;1) 
      =  -  \alpha  \dot{y} + m\ddot{y}$ &
   $\textstyle \frac{\mathcal{P}_{+}^{(1)}(\mathcal{W},t)} 
   {\tilde{\mathcal{P}}_{+}^{(1)}(-\mathcal{W},t)} 
   = \exp(\mathcal{W})$  \\
%
%&&\\
\hline
\end{tabular}
\vspfigC\vspace{0.5cm}
\end{table*}
%\end{center}
%\end{widetext}
%---------------------------------------------------------------------
%
   Now, we introduce the dimensionless modified ``work'' rate 
$\beta\Lambda_{\pm}(\ddot{y},\dot{y},y;\vartheta)v$ 
and its distribution function 
$\mathcal{P}_{\pm}^{(\vartheta)}(\mathcal{W},t)$ as 
\begin{eqnarray}
   \mathcal{P}_{\pm}^{(\vartheta)}(\mathcal{W},t) 
   = \pathaveA{\delta 
  \left(\mathcal{W}-\beta\int_{t_{0}}^{t} ds\; 
  \Lambda_{\pm}(\ddot{y},\dot{y},y;\vartheta)v\right)} 
\label{WorkDistr5}\end{eqnarray}
where $\pathaveA{\cdots}$ is the functional average 
in the inertia case, like the one given by Eq. (\ref{PathAvera1}). 
   Here, we remark that $\mathcal{W}$ in Eq. (\ref{WorkDistr5}) 
differs from the work 
$\mathcal{W}_{t}^{(v)}(\{y_{s}\})$ defined by Eq. (\ref{Work1}). 
   In a way similar to derive 
Eqs. (\ref{FluctTheorWork2}) and  
(\ref{FluctTheorFrict2}) in the over-damped case, 
it follows that the distribution function 
$\mathcal{P}_{\pm}^{(\vartheta)}(\mathcal{W},t)$ 
satisfies the transient fluctuation theorem 
\begin{eqnarray}
   \frac{\mathcal{P}_{\pm}^{(\vartheta)}(\mathcal{W},t)}
   {\tilde{\mathcal{P}}_{\pm}^{(\vartheta)}(-\mathcal{W},t)} 
   = \exp(\mathcal{W}) 
\label{FluctTheorModifWork1}\end{eqnarray}
under the condition that the initial distribution 
at time $t_{0}$ is given by 
the $f_{eq}^{(\vartheta)}(\dot{y},y)$.
   Here, the distribution 
$\tilde{\mathcal{P}}_{\pm}^{(\vartheta)}(\mathcal{W},t)$ 
is defined by 
\begin{eqnarray}
   &&\hspace{-0.5cm}
   \tilde{\mathcal{P}}_{\pm}^{(\vartheta)}(\mathcal{W},t) 
   \nonumber \\
   &&= \pathaveA{\delta 
   \left(\mathcal{W}-\beta\int_{t_{0}}^{t} ds\; 
   \Lambda_{\pm}(\ddot{y},\dot{y},y;
   (-1)^{\frac{1\pm 1}{2}}\vartheta)v\right)} ,
   \nonumber \\ 
\label{WorkDistr6}\end{eqnarray}
and is simply given by 
\begin{eqnarray}
   \tilde{\mathcal{P}}_{+}^{(0)}(\mathcal{W},t)
   &=&\mathcal{P}_{+}^{(0)}(\mathcal{W},t),  \\
    \tilde{\mathcal{P}}_{-}^{(\vartheta)}(\mathcal{W},t)
   &=&\mathcal{P}_{-}^{(\vartheta)}(\mathcal{W},t)
\label{WorkDistr7}\end{eqnarray}
in these special cases, and in order to derive 
Eq. (\ref{FluctTheorModifWork1}) 
we also used the relations 
$L^{(v)}\!\left(\ddot{y}_{s},\dot{y}_{s},y_{s}\right)
=L^{(-v)}\!\left(-\ddot{y}_{s},-\dot{y}_{s},-y_{s}\right)$, 
$\Lambda_{\pm}(\ddot{y}_{s},\dot{y}_{s},y_{s};\vartheta)
=-\Lambda_{\pm}(-\ddot{y}_{s},-\dot{y}_{s},-y_{s};\vartheta)$ 
and 
\begin{eqnarray}
   \Lambda_{\pm}(\ddot{y}_{s},\dot{y}_{s},y_{s};\vartheta) v 
   = - \Lambda_{\pm}(\ddot{y}_{s},-\dot{y}_{s},y_{s};
      (-1)^{\frac{1\pm 1}{2}}\vartheta) (\pm v) . 
      \nonumber \\
%\label{}
\end{eqnarray}
%. 
   It may be noted that the two terms $- \alpha  \dot{y}$
and $m\ddot{y}$ for the force 
$ \Lambda_{+}(\ddot{y}_{s},\dot{y}_{s},y_{s};1)$ 
have different time-reversal properties 
than the other forces 
$ \Lambda_{+}(\ddot{y}_{s},\dot{y}_{s},y_{s};0)$ and 
$ \Lambda_{+}(\ddot{y}_{s},\dot{y}_{s},y_{s};\vartheta)$. 

   From Eq. (\ref{FluctTheorModifWork1}) 
we derived four different fluctuation theorems 
corresponding to the cases 
$(\vartheta,\pm v) = (0,-v)$, $(1,-v)$, 
$(0,v)$, and $(1,v)$, where 
$\pm v$ is the velocity appearing in the Lagrangian function 
$L^{(\pm v)}(\ddot{y}_{s},-\dot{y}_{s},y_{s})$ 
on the right-hand side of the nonequilibrium detailed balance 
relation (\ref{DetaiBalan7}).  
   We summarize these four fluctuation theorems 
in Table \ref{4FluctTheor}. 
   In the last line of this table, appearance of the function  
$\tilde{\mathcal{P}}_{+}^{(1)}(-\mathcal{W},t)$ for the case 
of $(\vartheta,\pm v) = (1,v)$ is due to the different 
behavior with respect to time-reversal of the two terms 
$- \alpha \dot{y}$ and $m\ddot{y}$ 
composing the modified force 
$\Lambda_{+}(\ddot{y}_{s},\dot{y}_{s},y_{s};1)$, 
while in all the other cases in Table \ref{4FluctTheor} 
the modified forces have the unique behavior under 
time-reversal.

%%%%%%%%%%%%%%%%%%%%%%%%%%%%%%%%%%%%%%%%%%%%%%%%%%%%%%%%%%%%%%%%%%%%%%
\section{Conclusions and Remarks}
\label{ConclusionsRemarks}

   In this paper we discussed a generalization of 
Onsager-Machlup's fluctuation theory to 
nonequilibrium steady states and 
fluctuation theorems based on nonequilibrium 
detailed balance relations. 
   To that end, we used a model which consists 
of a Brownian particle confined by 
a harmonic potential which is dragged with a constant velocity 
through a heat reservoir. 
  This model is described by a Langevin equation, 
which is a simple 
and exactly-solvable nonequilibrium steady state model. 
   Our basic analytical approach   
is a functional integral technique, which was 
used in Onsager and Machlup's original work and 
is effective to discuss fluctuation theorems 
treating quantities expressed as functionals, 
for example, work and heat. 

   First, we gave an expression of the transition probability 
in terms of a Lagrangian function which can be written 
as a sum of an entropy production rate and two dissipation functions.  
   There is a difference, though with the similar result of 
Onsager and Machlup's original papers \cite{OM53,MO53},   
in that now the entropy production rate 
and one of the two dissipation functions 
- and consequently also the Lagrangian function - depend on 
the dragging velocity $v$ leading to nonequilibrium 
steady state effects.     
   From this property of the Lagrangian function,   
we constructed a nonequilibrium steady state thermodynamics 
by obtaining the second law of thermodynamics  
and the energy conservation law which involves  
fluctuating heat, work and an internal potential energy difference. 
   We also discussed Onsager's principle of 
minimum energy dissipation and the most probable path 
approximating the transition probability of the particle position. 
   This approach is different from another attempt 
for an Onsager-Machlup theory for nonequilibrium steady 
states \cite{BSG01,BSG02}, 
where a nonlinear diffusion equation is applied to models 
like an exclusion model and a boundary driven zero 
range model. 
   Instead, we use a stochastic model described 
by a Langevin equation, so that our results automatically include  
those of Onsager and Machlup's original works 
by taking a specific equilibrium value, $v=0$,  
for the nonequilibrium parameter $v$, 
%the velocity to drag the particle through the fluid.
and relax Onsager and Machlup's variable 
$\alpha$ and $\dot{\alpha}$ in Refs. \cite{OM53,MO53} 
to our variables 
$x$ and $\dot{x}$, respectively.   
  
   Second, we derived nonequilibrium detailed balance 
relations from the Lagrangian function  
to obtain not only the well-known fluctuation theorem for work 
but also another fluctuation theorem for energy loss by friction.  
   We also indicated the derivation of 
the extended fluctuation theorem 
for heat by carrying out explicitly the relevant functional integral 
and then using Refs. \cite{ZC03a,ZC04}.  
   In addition, we gave a simple argument for 
the heat fluctuation theorem in the long time limit.  
   Finally, we discussed briefly the effects of inertia, 
and obtained four different fluctuation theorems 
related to a potential force, a friction force 
and d'Alembert-like (or inertial) force 
for the comoving frame or the laboratory frame.  

   In the remaining of this section, we give remarks 
for the contents in the main text of this paper.    

   1) In this paper, we have emphasized a close connection 
between nonequilibrium detailed balance relations  
and fluctuation theorems, using a functional integral approach. 
   It may be noted that in some of past works 
concepts of detailed balances have been mentioned 
for formal derivations of fluctuation theorems 
in various different contexts, implicitly or explicitly 
\cite{ECM93,K98,LS99,C99,C00,CCJ06}. 
   However, we should keep in mind that 
a generalization of the equilibrium detailed balance 
to nonequilibrium states is not unique, as shown in this paper 
[cf. Eq. (\ref{DetaiBalan1}) and (\ref{DetaiBalan3})].  
   As a remark related to this point, 
we should notice that 
even if the equilibrium detailed balance condition 
is violated, but another detailed balance condition 
for the nonequilibrium steady state still holds, 
namely, using the nonequilibrium steady state distribution 
$f_{ss}(y)$ we obtain for the over-damped case: 
\begin{eqnarray}
   e^{\int_{t_{0}}^{t}ds\; L^{(v)}\!
      \left(\dot{y}_{s},y_{s}\right)}
      f_{ss}(y_{0}) 
   = f_{ss}(y_{t}) 
      \;e^{\int_{t_{0}}^{t}ds\; L^{(v)}\!
      \left(-\dot{y}_{s},y_{s}\right)} , \;\;\;
\label{DetaiBalan5}\end{eqnarray}
by Eqs. (\ref{Lagra1}), (\ref{SteadSolut1}) 
and (\ref{EquilDistr1}), 
or equivalently  
$\transS{F}{y_{t}}{t}{y_{0}}{t_{0}} f_{ss}(y_{0}) 
   = \transS{F}{y_{0}}{t}{y_{t}}{t_{0}} f_{ss}(y_{t})$.  
   Here, it is essential to note that on the right-hand side 
of Eq. (\ref{DetaiBalan5}) we do not change the sign of 
the dragging velocity $v$ although we change the sign 
of the particle velocity $\dot{y}_{s}$ in the comoving frame. 
   We emphasize that here, there is no additional multiplying factors 
like $\exp[-\beta\mathcal{W}_{t}^{(v)}(\{y_{s}\})]$ as in 
Eq. (\ref{DetaiBalan1}) or 
$\exp[-\beta\mathcal{R}_{t}^{(v)}(y_{t},y_{0})]$ 
as in Eq. (\ref{DetaiBalan3}). 
   As a consequence we have been unable to derive 
fluctuation theorems from Eq. (\ref{DetaiBalan5}). 
%which are vital to derive fluctuation theorems.  
   Since we chose the equilibrium state 
as the reference state for the detailed balance 
in this paper,   
our interest was mainly the work 
to maintain the system in a nonequilibrium steady state,  
i.e. the work necessary to keep the system from going to   
the equilibrium state. 
   In this sense, we note that the reference state is arbitrary,  
for example, if we are interested in the work to go  
from one nonequilibrium state to another nonequilibrium state. 
   In general, the modification of the detailed balance relation 
based on an arbitrary reference distribution function $f_{ref}(y)$ 
can be expressed as 
\begin{eqnarray}
   && 
   e^{-\beta Y_{t}(\{y_{s}\})}
      e^{\int_{t_{0}}^{t}ds\; L^{(v)}\!
      \left(\dot{y}_{s},y_{s}\right)}
      f_{ref}(y_{0}) 
      \nonumber \\
   && \spaEq 
   = f_{ref}(y_{t}) 
      \;e^{\int_{t_{0}}^{t}ds\; L^{(-v)}\!
      \left(-\dot{y}_{s},y_{s}\right)} 
\label{DetaiBalan6}\end{eqnarray}
where $Y_{t}(\{y_{s}\})$ is 
the functional defined by 
\begin{eqnarray}
   Y_{t}(\{y_{s}\}) \equiv \mathcal{Q}^{(v)}_{t}(\{y_{s}\}) 
   +\beta^{-1}\ln\frac{f_{ref}(y_{0})}{f_{ref}(y_{t})} .
%\label{}
\end{eqnarray}
   Choosing $f_{ref}(y)=f_{eq}(y)$, Eq. 
(\ref{DetaiBalan6}) leads to Eq. (\ref{DetaiBalan1}).     
   We can also get 
%a relation similar to Eq. (\ref{DetaiBalan6}) 
a generalization of Eq. (\ref{DetaiBalan3}) 
for an arbitrary reference distribution function $f_{ref}(y)$.  
%by changing $-v$ on the right-hand 
%of Eq. (\ref{DetaiBalan6}) to $v$. 
   An analogous quantity to 
$Y_{t}(\{y_{s}\})$ is in Ref. \cite{ES02} 
for a thermostated system with deterministic dynamics. 

   2) From Eq. (\ref{DetaiBalan5}) we derive 
\begin{eqnarray}
   &&
  \int_{t_{0}}^{t}ds\; L^{(v)}\!
      \left(\dot{y}_{s},y_{s}\right) 
      + \tilde{S}_{ss}(y_{0})/k_{B} 
      \nonumber \\
   &&\spaEq 
   = \int_{t_{0}}^{t}ds\; L^{(v)}\!
      \left(-\dot{y}_{s},y_{s}\right) 
      + \tilde{S}_{ss}(y_{t})/k_{B} 
\label{OMSymme1}\end{eqnarray}
where $\tilde{S}_{ss}(y)$ is defined by  
%
%\begin{eqnarray}
$   \tilde{S}_{ss}(y) \equiv k_{B}\ln f_{ss}(y) $.
%\label{EntroStati1}\end{eqnarray}
%
   An identity like Eq. (\ref{OMSymme1}) is called 
an Onsager-Machlup symmetry \cite{BSG01,BSG02,G02}, 
and is used to discuss 
macroscopic properties of nonequilibrium steady states. 
   Using Eq. (\ref{OMSymme1}) we can also obtain 
an expression like 
\begin{eqnarray}
   \frac{\exp\left[\int_{t_{0}}^{t}ds\; L^{(v)}\!
      \left(\dot{y}_{s},y_{s}\right)\right]}
      {\exp\left[\int_{t_{0}}^{t}ds\; L^{(v)}\!
      \left(-\dot{y}_{s},y_{s}\right)\right]} 
      = \exp\left[\beta \tilde{Q}_{ss}(t,t_{0})\right]
\label{OMSymme2}\end{eqnarray}
with  
$\tilde{Q}_{ss}(t,t_{0}) \equiv T [ \tilde{S}_{ss}(y_{t}) 
- \tilde{S}_{ss}(y_{0}) ]$. 
   On the other hand, it can be shown 
from Eqs. (\ref{EquilDistr1}), (\ref{EnergyBalan1}) 
and (\ref{DetaiBalan1}) 
[or from Eq. (\ref{DetaiBalan6})] that 
\begin{eqnarray}
   \frac{\exp\left[\int_{t_{0}}^{t}ds\; L^{(v)}\!
      \left(\dot{y}_{s},y_{s}\right)\right]}
      {\exp\left[\int_{t_{0}}^{t}ds\; L^{(-v)}\!
      \left(-\dot{y}_{s},y_{s}\right)\right]} 
      = \exp\left[\beta \mathcal{Q}^{(v)}_{t}(\{y_{s}\})\right]
\label{OMSymme3}\end{eqnarray}
using the heat $\mathcal{Q}^{(v)}_{t}(\{y_{s}\})$ 
of Eq. (\ref{Heat1}). 
%discussed in Eq. (\ref{EnergyBalan1}). 
%   [Note the difference of the sign of $v$ in 
%the denominators on the left-hand sides of Eqs. (\ref{OMSymme2}) 
%and (\ref{OMSymme3}).] 
   Note that Eq. (\ref{OMSymme3}) 
is consistent with the heat $\mathcal{Q}^{(v)}_{t}(\{y_{s}\})$ 
appearing in our energy conservation law 
(\ref{EnergyBalan1}), 
in contrast to Eq. (\ref{OMSymme2}) 
in which the quantity $\tilde{Q}_{ss}(t,t_{0})$ does not have  
such a correspondence with the heat. 
   Thus we will restrict ourselves in the following 
to Eq. (\ref{OMSymme3}). 
%although some of the comments below 
%can also be for Eq. (\ref{OMSymme2}) similarly.
% 
   As we discussed in Sec. \ref{DraggedBrownianParticle}, 
the term $\exp[\int_{t_{0}}^{t}ds\; L^{(v)}\!
(\dot{y}_{s},y_{s})]$ 
appearing in the numerator on the left-hand side 
of Eq. %(\ref{OMSymme2}) and 
(\ref{OMSymme3}) is 
the probability functional of the forward path 
$\{y_{s}\}_{s\in[t_{0},t]}$. 
%[or a transition probability 
%if we use the most probable path 
%for the path $\{y_{s}\}_{s\in[t_{0},t]}$ in the sense of 
%Eq. (\ref{OMformu1})]. 
   On the other hand, the denominator on the left-hand side  
of Eq. %(\ref{OMSymme2}) and 
(\ref{OMSymme3}) is  
the probability functional of the corresponding 
time-reversed path with the dragging velocity 
%$v$ or 
$-v$. 
%, respectively 
%[or a backward transition probability 
%when we use the time-reversed most probable path]. 
   Therefore, Eq. %(\ref{OMSymme2}) and 
(\ref{OMSymme3}) implies that 
the logarithm of the ratio of such forward and backward 
probability functionals is given by 
the heat %(or a heat-like quantity) 
multiplied by the inverse temperature.
   In this sense it is tempting to claim 
Eq. (\ref{OMSymme3}) as a relation leading  
to a fluctuation theorem \cite{ND04,K05,IP06}. 
   However, it is important to distinguish 
Eq. %(\ref{OMSymme2}) and 
(\ref{OMSymme3}) from the  
fluctuation theorems discussed in the main text of this paper.  
   First of all, although it is related to the heat,  
Eq. %(\ref{OMSymme2}) and 
(\ref{OMSymme3}) has a form rather close to a relation 
leading to the conventional fluctuation theorems 
like Eqs. (\ref{FluctTheorWork2}) and (\ref{FluctTheorFrict2}), 
which are different from the extended form for 
the heat fluctuation theorem 
discussed in Sec. \ref{FluctuationTheoremHeat}. 
   Similarly, although one may regard Eq. (\ref{OMSymme3}) 
as a nonequilibrium detailed balance relation for heat, 
a derivation of the extended fluctuation theorem for heat 
from a nonequilibrium detailed balance relation 
remains an open problem.   
   We should also notice that 
no initial condition dependence appears in 
Eq. %(\ref{OMSymme2}) and 
(\ref{OMSymme3}), so that 
we cannot discuss directly, for example, a difference 
between the transient fluctuation theorem 
and the steady state fluctuation theorem 
for them from Eq. %(\ref{OMSymme2}) and 
(\ref{OMSymme3}). 

   3) Although the nonequilibrium detailed balance relations,  
like Eq. (\ref{DetaiBalan1}), (\ref{DetaiBalan3}) 
or (\ref{DetaiBalan6}), play an essential role to derive 
the fluctuation theorem, it is important to note that some 
properties of the fluctuation theorem cannot 
be discussed by it. 
   Basically, the nonequilibrium detailed balance relation 
can lead directly to the so-called transient fluctuation theorems, 
which are identically satisfied for any time \cite{CG99}, 
but this relation does not say 
what happens to fluctuation theorems  
%for a specific quantity (e.g. work and heat) 
if we change the initial condition 
(like the equilibrium distribution) 
to another (like the nonequilibrium steady state 
discussed in the steady state fluctuation theorem). 
   The transient fluctuation theorem can be 
different from the steady state fluctuation theorem 
for some quantities, even in the long time limit. 
   As an example for such a difference,  
we showed in this paper 
that the energy loss by friction satisfies the 
transient fluctuation theorem but does not satisfy 
the steady state fluctuation theorem.  

   We have discussed an initial condition dependence 
of fluctuation theorems by carrying out functional integrals 
to obtain distribution functions explicitly, 
and showed that the work distribution function 
has an asymptotic form satisfying the work fluctuation theorem, 
independent of the initial distribution, 
while the friction-loss distribution function 
depends on the initial condition even in the long time limit. 
   This difference between the work and the friction-loss 
might come from the fact that the work 
is given by a time-integral of the particle position 
so that its contribution near the initial time can be 
neglected in the long time limit, while the energy loss by friction  
is given by the particle position at the initial 
and final times only.    
   A systematic way to investigate whether a fluctuation 
theorem is satisfied for any initial condition 
without calculating a distribution function,   
remains an open problem \cite{noteHeat}. 

   4) Finally we note that the analogy of the Brownian 
particle case, discussed here, and the electric circuit case 
should persist not only in the over-damped case 
(as shown in \cite{ZCC04}) but also in the case 
including inertia.
   In that case, one has to 
add the self-induction $L_{0}$ of the electric circuit, 
as the corresponding quantity of the mass $m$ 
of the Brownian particle. 
   This will add the correspondence of $m$ and $L_{0}$  
to Table I in Ref. \cite{ZCC04}.
   Then, the fluctuation theorems in Table \ref{4FluctTheor} 
in Sec. \ref{InertiaEffects} of this paper can, 
by using the extended analogy described above, 
also be used for electric circuits, 
and might be experimentally accessible (cf. \cite{GC05}). 

%   In this paper we have concentrated on a specific model 
%of a dragged Brownian particle to discuss Onsager-Machlup theory 
%and fluctuation theorems for nonequilibrium steady states. 
%   We chose our discussions in this way, in order to 
%make our discussion as simple and concrete as possible, 
%which may be helpful to clarify physical meanings 
%of our arguments and results.  
%   On the other hand, many generalizations of our work  
%can be considered, for example, to more general potential 
%form, a time-dependent dragging velocity, cases of many particles, 
%non-Gaussian cases, nonlinear cases, etc. 
%   It may also be interesting to compare our arguments using 
%the functional integral approach with the ones using 
%other stochastic approaches, for example, 
% using a Fokker-Planck equation \cite{H77} instead of 
%Onsager-Machlup's functional approach, and 
%the energy conservation law using a Langevin equation approach 
%\cite{S98}, etc. 
%   These remain as future problems.  

%%%%%%%%%%%%%%%%%%%%%%%%%%%%%%%%%%%%%%%%%%%%%%%%%%%%%%%%%%%%%%%%%%%%%%
\section*{Acknowledgements}

   One of the authors (E.G.D.C.) is indebted to Prof. 
G. Jona-Lasinio for a stimulating discussion at the Isaac Newton 
Institute for Mathematical Sciences in Cambridge, U.K, 
while the other author (T.T.) wishes to express his gratitude 
to Prof. K. Kitahara for a comment about a relation 
between the Onsager-Machlup theory and 
a fluctuation theorem near equilibrium.
   Both authors gratefully acknowledge financial support 
of the National Science Foundation, under award PHY-0501315.

%%%%%%%%%%%%%%%%%%%%%%%%%%%%%%%%%%%%%%%%%%%%%%%%%%%%%%%%%%%%%%%%%%%%%%
%\pagebreak
\appendix
%\setcounter{section}{0} 
%\makeatletter 
%   \@addtoreset{equation}{section} 
%   \makeatother 
%   \def\theequation{\Alph{section}.% 
%   \arabic{equation}} 
   
\section{Transition Probability using a Functional Integral Technique} 
\label{TransitionProbabilityFunctionalIntegral}

   In this Appendix, we outline a derivation of the 
transition probability (\ref{TransProba1}) for the stochastic 
process described by the Langevin equation (\ref{LangeEq2}). 

   First, we translate the Langevin equation (\ref{LangeEq2}) 
into the corresponding Fokker-Planck equation. 
   It can be done using the Kramers-Moyal expansion technique 
\cite{K92,R89}, and we obtain the Fokker-Planck equation 
\begin{eqnarray}
   \frac{\partial f(y,t)}{\partial t} 
   = \hat{\mathcal{L}} f(y,t) 
\label{FPequat1}\end{eqnarray} 
for the distribution function $f(y,t)$ of the particle 
position $y$ at time $t$. 
   Here, $\hat{\mathcal{L}}$ is the Fokker-Planck 
operator defined by 
\begin{eqnarray}
   \hat{\mathcal{L}} \equiv
   \frac{\partial}{\partial y} 
   \left( \frac{y + v\tau}{\tau} + D
   \frac{\partial}{\partial y} \right) 
\label{FPopera1}\end{eqnarray} 
with $D\equiv 1/(\alpha\beta)$. 

   The transition probability 
$\transS{F}{y}{t+\Delta t}{y'}{t}$ 
from $y'$ at time $t$ to $y$ at time $t+\Delta t$ is given by 
\begin{widetext}
\begin{eqnarray}
%c   &&
   \transP{F}{y}{t +\Delta t}{y'}{t} 
%c   \nonumber \\
%c   &&\spaEq 
%   =
   &=& 
   e^{\hat{\mathcal{L}} \Delta t}\delta (y-y')
   \nonumber \\
%   &&\spaEq = 
   &=& 
      \left[ 1 +  \hat{\mathcal{L}} \Delta t 
      + \mathcal{O}\left(\Delta t^{2}\right)
      \right] 
      \delta \left(y-y'\right)
      \nonumber \\
%   &&\spaEq = 
   &=& 
      \left( 1 + \Delta t 
       \frac{\partial}{\partial y} 
      \frac{y'+v\tau}{\tau} 
      +D\Delta t \frac{\partial^{2}}{\partial y^{2}} 
      \right) 
      \delta \left(y-y'\right)
      + \mathcal{O}\left(\Delta t^{2}\right)
      \nonumber \\
   &=&
      \frac{1}{2\pi} \int_{-\infty}^{+\infty} d\lambda \; 
      \left( 1 + \Delta t 
      \frac{y'+v\tau}{\tau}  
       \frac{\partial}{\partial y} 
      +D\Delta t \frac{\partial^{2}}{\partial y^{2}} 
      \right) 
%c   \nonumber \\
%c   &&\spaEq\spaEq \times 
      \exp\left[ i \lambda (y-y')
      \right]
 %     \nonumber \\
%   &&\spaEq      
      + \mathcal{O}\left(\Delta t^{2}\right)
      \nonumber \\
   &=& 
      \frac{1}{2\pi}\int_{-\infty}^{+\infty} d\lambda \;
      \exp\left[-D\Delta t  
      \lambda^{2} + i \Delta t  
      \left( \frac{y-y'}{\Delta t } + 
      \frac{y'+v\tau}{\tau}  \right) \lambda \right]
 %     \nonumber \\
 %  &&\spaEq
      + \mathcal{O}\left(\Delta t^{2}\right)
      \nonumber \\
%   &&\spaEq = 
   &=&
   \frac{1}{\sqrt{4\pi D\Delta t}} 
      \exp\left[- \frac{1}{4D}
      \left(\frac{y-y'}{\Delta t} + \frac{y'+v\tau}{\tau} \right)^{2}
       \Delta t \right] 
%c   \nonumber \\
%c   &&\spaEq\spaEq 
   + \mathcal{O}(\Delta t^{2}) 
\label{TransProba3}\end{eqnarray}
%\end{widetext}
%
where we used $y \delta \left(y-y'\right) 
= y' \delta \left(y-y'\right)$.
   On the other hand, using the Chapman-Kolmogorov equation \cite{K92}, 
the transition probability for a finite time interval $t-t_{0}$ 
is expressed as  
\begin{eqnarray}
   \transP{F}{y_{t}}{t}{y_{0}}{t_{0}} &=& 
      \lim_{N\rightarrow+\infty} 
      \int dy_{N-1} \int dy_{N-2} \cdots \int dy_{1} \; 
    %\nonumber \\  
   %&& \spaEq \times 
      \transP{F}{y_{t}}{t}{y_{N-1}}{t_{N-1}}
      \transP{F}{y_{N-1}}{t_{N-1}}{y_{N-2}}{t_{N-2}}
  %    \nonumber \\  
  % && \spaEq 
   \cdots  \transP{F}{y_{1}}{t_{1}}{y_{0}}{t_{0}} 
\label{TransProba4}\end{eqnarray}
\end{widetext}
with $t_{n}\equiv t_{0} + n \Delta t_{N}$, $n=1,2,\cdots,N$,  
$\Delta t_{N}\equiv (t-t_{0})/N$, $t_{N}=t$. 
   Inserting the expression (\ref{TransProba3}) for the transition 
probability in a short time interval $\Delta t = \Delta t_{N}$ into 
Eq. (\ref{TransProba4}) we obtain Eq. (\ref{TransProba1}) 
with the functional integral (\ref{FunctInteg1}).

%   or simply solving the Fokker-Planck equation (\ref{FPequat1})

%%%%%%%%%%%%%%%%%%%%%%%%%%%%%%%%%%%%%%%%%%%%%%%%%%%%%%%%%%%%%%%%%%%%%%
\section{Fluctuation Theorem for Work}
\label{TransientFluctuationTheoremWork}

   In this Appendix, we show the relation 
$\mathcal{E}_{w}^{(v)}(\lambda,t) 
= \mathcal{E}_{w}^{(-v)}(1-\lambda,t)$, therefore 
Eq. (\ref{FluctTheorWork2}). 
   We also give a derivation of Eq. (\ref{FluctTheorWork2}) 
from Eq. (\ref{FluctTheorWork1}). 

%---------------------------------------------------------------------
   From Eq. (\ref{EFunctWork1}) with the functional average 
(\ref{PathAvera1}) we derive 
\begin{widetext}
\begin{eqnarray}
   \mathcal{E}_{w}^{(v)}(\lambda,t) 
%   &=&
%   \pathaveA{e^{-\lambda\beta 
%   \mathcal{W}_{t}^{(v)}(\{y_{s}\})}}
%   \nonumber \\
   &=& \int dy_{t} \! 
      \int_{y_{0}}^{y_{t}}\mathcal{D}y_{s} \!\int dy_{0} \;   
      e^{\int_{t_{0}}^{t}ds\; 
      L^{(v)}\!\left(\dot{y}_{s},y_{s}\right)}
      f(y_{0},t_{0}) 
      \;e^{-\lambda \beta 
      \mathcal{W}_{t}^{(v)}(\{y_{s}\})}
   \nonumber \\
   &=& \int dy_{t} \! 
      \int_{y_{0}}^{y_{t}}\mathcal{D}y_{s} \!\int dy_{0} \;   
      f_{eq}(y_{t}) 
      \;e^{\int_{t_{0}}^{t}ds\; L^{(-v)}\!
      \left(-\dot{y}_{s},y_{s}\right)}
      \;e^{\beta\mathcal{W}_{t}^{(v)}(\{y_{s}\})}
 %  \nonumber \\
 %  && \hspace{2cm} \times
      \frac{1}{f_{eq}(y_{0})} \; f(y_{0},t_{0})
      \;e^{-\lambda \beta 
      \mathcal{W}_{t}^{(v)}(\{y_{s}\})}
   \nonumber \\
%   &=& \int dy_{t} \! 
%      \int_{y_{0}}^{y_{t}}\mathcal{D}y_{s} \!\int dy_{0} \;   
%      f_{eq}(y_{t}) 
%      \;e^{\int_{t_{0}}^{t}ds\; L^{(-v)}\!
%      \left(-\dot{y}_{s},y_{s}\right)}
%      \;e^{(1-\lambda)\beta\mathcal{W}_{t}^{(v)}(\{y_{s}\})}
%   \nonumber \\
   &=&  \int dy_{0} \! 
      \int_{y_{0}}^{y_{t}}\mathcal{D}y_{s} \! \int dy_{t} \;   
      e^{\int_{t_{0}}^{t}ds\; L^{(-v)}\!
      \left(-\dot{y}_{s},y_{s}\right)}
      f_{eq}(y_{t}) 
      \;e^{-(1-\lambda)\beta\mathcal{W}_{t}^{(-v)}(\{y_{s}\})}
   \label{FunctE2b} \\
   &=& \int dy_{t} \! 
      \int_{y_{0}}^{y_{t}}\mathcal{D}y_{s} \!\int dy_{0} \;   
      e^{\int_{t_{0}}^{t}ds\; L^{(-v)}\!
      \left(\dot{y}_{s},y_{s}\right)}
      f_{eq}(y_{0}) 
      \;e^{-(1-\lambda)\beta\mathcal{W}_{t}^{(-v)}(\{y_{s}\})}
   \label{FunctE2c} \\
   &=& \mathcal{E}_{w}^{(-v)}(1-\lambda,t)
\label{FluctTheor2}\end{eqnarray}
\end{widetext}
where we used Eqs. (\ref{DetaiBalan1})   
%(\ref{DimenLessWorkRate}) 
and $\mathcal{W}_{t}^{(-v)}(\{y_{s}\}) 
= - \mathcal{W}_{t}^{(v)}(\{y_{s}\})$, 
and the assumption 
$f(y_{0},t_{0})=f_{eq}(y_{0})$. 
   Here, in the transformation from Eq. (\ref{FunctE2b}) to 
Eq. (\ref{FunctE2c}) we changed the integral variables 
as $y_{s}\rightarrow y_{t+t_{0}-s}$ 
(so that $\dot{y}_{s}\rightarrow -\dot{y}_{s}$, 
$y_{t}\rightarrow y_{0}$ 
and $y_{0}\rightarrow y_{t}$). 
   Therefore, we obtain $\mathcal{E}_{w}^{(v)}(\lambda,t) 
= \mathcal{E}_{w}^{(-v)}(1-\lambda,t)$, 
whose combination with Eq (\ref{EFunctWorkRelat1}) leads to  
Eq. (\ref{FluctTheorWork1}).

%---------------------------------------------------------------------
   Moreover, from Eqs. (\ref{WorkDistr2}) and 
(\ref{FluctTheorWork1}) we derive  
\begin{eqnarray}
   P_{w}(W,t) 
%   &=& \frac{1}{2\pi} 
%      \int_{-\infty}^{+\infty}d\lambda \;
%      e^{i\lambda W} \mathcal{E}_{w}^{(v)}(i\lambda)
%   \nonumber \\
   &=& \frac{1}{2\pi}\int_{-\infty}^{+\infty}d\lambda \;
      e^{i\lambda W} \mathcal{E}_{w}^{(v)}(1-i\lambda)
   \nonumber \\
   &=& \frac{1}{2\pi}\int_{-\infty-i}^{+\infty-i}d\mu \;
      e^{(1-i\mu) W} \mathcal{E}_{w}^{(v)}(i\mu)
   \label{WorkRate3a} \\
   &=& e^{W} \frac{1}{2\pi} \int_{-\infty}^{+\infty}d\mu \;
      e^{i \mu (-W)} \mathcal{E}_{w}^{(v)}(i\mu) 
   \label{WorkRate3b} \\
   &=& e^{W} P_{w}(-W,t)
\label{WorkRate3d}\end{eqnarray}
with $\mu \equiv - \lambda - i$.
   Here, in the transformation from Eq. (\ref{WorkRate3a}) 
to Eq. (\ref{WorkRate3b}) we used the fact that 
noting Eq. (\ref{EFunctWork1}) the function 
$\mathcal{E}_{w}^{(v)}(i\mu)\exp[(1-i\mu) W]$ 
appearing in Eq. (\ref{WorkRate3a}) does not 
have any pole in the complex region for $\mbox{Im}\{\mu\}\in [0,1]$ 
(here $\mbox{Im}\{\mu\}$ is the imaginary part of $\mu$). 
   Using Eq. (\ref{WorkRate3d}) we obtain 
Eq. (\ref{FluctTheorWork2}).

%%%%%%%%%%%%%%%%%%%%%%%%%%%%%%%%%%%%%%%%%%%%%%%%%%%%%%%%%%%%%%%%%%%%%%
\section{Functional Integral Calculation for the Work Distribution}
\label{CalculationWorkDistribution}

   In this Appendix, we give calculation details of 
Eqs. (\ref{EularLagra2Solve1}), 
(\ref{EFunctWorkTrnas2}) and (\ref{PathInteg1}). 
%and (\ref{EFunctWork3}).

%---------------------------------------------------------------------
%\subsection{Derivation of Eq.  (\ref{EularLagra2Solve1})}

   Inserting Eq. (\ref{WorkRate1}) and (\ref{Lagra1}) into 
Eq. (\ref{EularLagra2Modif1}), we obtain 
\begin{eqnarray}
      \frac{d^{2} \ym_{s}}{ds^{2}} 
         = \frac{\ym_{s} + (1-2\lambda) v\tau}{\tau^{2}}  
\label{EularLagraModif1}\end{eqnarray}
where we used the relations $\alpha = \kappa\tau$ 
and $D=1/(\alpha\beta)$.
   [Note that Eq. (\ref{EularLagraModif1}) 
for $\ym_{s}$ is Eq. (\ref{EularLagra2}) for $y_{s}^{*}$ 
except for that Eq. (\ref{EularLagraModif1}) 
use $(1-2\lambda) v$ instead of $v$ in Eq. (\ref{EularLagra2}).] 
   The solution of Eq. (\ref{EularLagraModif1}) is given by 
\begin{eqnarray} 
   \ym_{s} +(1-2\lambda) v\tau 
   = \tilde{A}_{1}\exp\left(\frac{s}{\tau}\right) 
      + \tilde{A}_{2}\exp\left(-\frac{s}{\tau}\right) .
\label{SolutLagra1}\end{eqnarray}
   Here, $\tilde{A}_{1}$ and $\tilde{A}_{2}$ are constants 
determined by the conditions  
$\ym_{t}=y_{t}$ and $\ym_{0}(=\ym_{t_{0}})=y_{0}$, namely 
\begin{eqnarray}
   &&\left(\begin{array}{c} 
      y_{0} + (1-2\lambda)v\tau  \\
      y_{t} + (1-2\lambda)v\tau
      \end{array}\right)
      \nonumber\\ 
   &&\spaEq = 
   \left(\begin{array}{cc}
      \exp\left(\frac{t_{0}}{\tau}\right)  & 
      \exp\left(-\frac{t_{0}}{\tau}\right) \\
      \exp\left( \frac{t}{\tau}\right) & 
      \exp\left(-\frac{t}{\tau}\right) 
      \end{array}\right)
   \left(\begin{array}{c} 
      \tilde{A}_{1}  \\
      \tilde{A}_{2} 
      \end{array}\right) . 
\label{SolutLagraConst1}\end{eqnarray}
   Solving Eq. (\ref{SolutLagraConst1}) for 
$\tilde{A}_{1}$ and $\tilde{A}_{2}$ we obtain 
\begin{eqnarray}
   \left(\begin{array}{c} 
      \tilde{A}_{1}  \\
      \tilde{A}_{2} 
      \end{array}\right) 
   =\left(\begin{array}{c} 
      A_{t-t_{0}}^{((1-2\lambda)v)}(y_{t},y_{0})
      \exp\left(-\frac{t}{\tau}\right) \\
      A_{-(t-t_{0})}^{((1-2\lambda)v)}(y_{t},y_{0}) 
      \exp\left(\frac{t}{\tau}\right) 
      \end{array}\right)
\label{SolutLagraConst2}\end{eqnarray}
with the function $A_{t-t_{0}}^{(v)}(y_{t},y_{0})$ defined by 
Eq.  (\ref{ConstA1}). 
   Further, we note 
\begin{eqnarray}
   A_{-(t-t_{0})}^{(v)}(y_{t},y_{0}) 
   = A_{t-t_{0}}^{(v)}(y_{0},y_{t}) 
   b_{t} .
\label{SolutLagraConst3}\end{eqnarray}
which can be shown from Eq.  (\ref{ConstA1}).
   Using Eqs. (\ref{SolutLagra1}), (\ref{SolutLagraConst2}) 
and (\ref{SolutLagraConst3}) we obtain Eq.  
(\ref{EularLagra2Solve1}).

%---------------------------------------------------------------------
   Noting that the Lagrangian function 
$L^{(v)}(\dot{y}_{s},y_{s})$ 
defined by Eq. (\ref{Lagra1}) 
and the work rate $\dot{\mathcal{W}}^{(v)}(y_{t})$ 
given by Eq. (\ref{WorkRate1}) are the second order to 
$y_{t}$ and $\dot{y}_{s}$ at most, we obtain 
\begin{widetext}
\begin{eqnarray}
   &&\int_{t_{0}}^{t} ds
      \left[L^{(v)}(\dot{y}_{s},y_{s}) 
      -\lambda \beta\dot{W}(y_{s})\right]
      \nonumber \\
   &&\spaEq = \int_{t_{0}}^{t} ds 
      \left[L^{(v)}(\dot{\ym}_{s}
         +\dot{\tilde{z}}_{s},\ym_{s}+\tilde{z}_{s}) 
      -\lambda \beta\dot{W}(\ym_{s}+\tilde{z}_{s})\right]
      \nonumber \\ 
   &&\spaEq = \int_{t_{0}}^{t} ds \; \left\{
      L^{(v)}(\dot{\ym}_{s},\ym_{s}) 
      -\lambda \beta\dot{W}(\ym_{s})
      +\frac{\partial 
      \left[L^{(v)}(\dot{\ym}_{s},\ym_{s}) 
      -\lambda \beta\dot{W}(\ym_{s})\right]} 
         {\partial \dot{\ym}_{s}} \dot{\tilde{z}}_{s}
      \right. \nonumber \\ 
   &&\spaEq\spaEq\spaEq \left.
      +\frac{\partial 
      \left[L^{(v)}(\dot{\ym}_{s},\ym_{s}) 
      -\lambda \beta\dot{W}(\ym_{s})\right]} 
         {\partial \ym_{s}} \tilde{z}_{s}
      +\frac{1}{2}\frac{\partial^{2} 
      \left[L^{(v)}(\dot{\ym}_{s},\ym_{s}) 
      -\lambda \beta\dot{W}(\ym_{s})\right] } 
         {\partial \dot{\ym}_{s}{}^{2}} 
         \dot{\tilde{z}}_{s}^{2}
      \right. \nonumber \\ 
   &&\spaEq\spaEq\spaEq \left.
      +\frac{\partial^{2} 
      \left[L^{(v)}(\dot{\ym}_{s},\ym_{s}) 
      -\lambda \beta\dot{W}(\ym_{s})\right] } 
         {\partial \dot{\ym}_{s}
            \partial \ym_{s}} 
            \dot{\tilde{z}}_{s}\tilde{z}_{s}
      +\frac{1}{2}\frac{\partial^{2} 
      \left[L^{(v)}(\dot{\ym}_{s},\ym_{s}) 
      -\lambda \beta\dot{W}(\ym_{s})\right] } 
         {\partial \ym_{s}{}^{2}} \tilde{z}_{s}^{2}
      \right\}
      \nonumber \\ 
   &&\spaEq = \int_{t_{0}}^{t} ds \; \Biggl\{
      L^{(v)}(\dot{\ym}_{s},\ym_{s}) 
      -\lambda \beta\dot{W}(\ym_{s}) %\right. 
%      \nonumber \\ 
%   &&\spaEq\spaEq\spaEq 
      \left.
      - \left[
      \frac{d}{ds}
      \frac{\partial 
      L^{(v)}(\dot{\ym}_{s},\ym_{s})} 
         {\partial \dot{\ym}_{s}} 
      -\frac{\partial L^{(v)}(\dot{\ym}_{s},\ym_{s})} 
         {\partial \ym_{s}} 
      +\lambda \beta\frac{\partial \dot{W}(\ym_{s})} 
         {\partial \ym_{s}} 
      \right] \tilde{z}_{s}
      \right. \nonumber \\ 
   &&\spaEq\spaEq\spaEq %\left.
      -\frac{1}{4D}\left(
       \dot{\tilde{z}}_{s}^{2}
      +\frac{2}{\tau} \dot{\tilde{z}}_{s}\tilde{z}_{s}
      +\frac{1}{\tau^{2}} \tilde{z}_{s}^{2}
      \right)\Biggr\}
      \nonumber \\ 
%   &&\spaEq =  \int_{t_{0}}^{t} ds \; \left[
%      L^{(v)}(\dot{\ym}_{s},\ym_{s}) 
%      -\lambda \beta\dot{W}(\ym_{s})
%      -\frac{1}{4D}\left(
%       \dot{\tilde{z}}_{s}
%      +\frac{1}{\tau} \tilde{z}_{s}
%      \right)^{2}\right]
%      \nonumber \\ 
   &&\spaEq = \int_{t_{0}}^{t} ds \; \left[
      L^{(v)}(\dot{\ym}_{s},\ym_{s}) 
      -\lambda \beta\dot{W}(\ym_{s})
      +L^{(0)}(\dot{\tilde{z}}_{s},\tilde{z}_{s}) 
      \right]
\label{ActioDecom1}\end{eqnarray}
\end{widetext}
using 
%the relation $y_{s}=\ym_{s}+\tilde{z}_{s}$, 
a partial integral and 
%the Eular-Lagrange equation 
Eqs. (\ref{Lagra1}), (\ref{EularLagra2Modif1}) 
and (\ref{VariaZ1}). 
   Inserting Eq. (\ref{ActioDecom1}) into 
Eq. (\ref{EFunctWorkTrnas1})
we obtain Eq. (\ref{EFunctWorkTrnas2}).

%---------------------------------------------------------------------
   Noting $t_{n}\equiv t_{0} + n \Delta t_{N}$, $n=1,2,\cdots,N$,  
$\Delta t_{N}\equiv (t-t_{0})/N$,  
the initial time $t_{0}$, the final time $t_{N}=t$, 
and $\tilde{z}_{0}\equiv \tilde{z}_{t_{0}}=0$ 
from Eq. (\ref{VariaZBound1}) we have   
\begin{eqnarray}
   &&
   \sum_{n=0}^{k}\left( 
      \varphi \tilde{z}_{t_{n}} + \tilde{z}_{t_{n+1}}\right)^{2} 
      \nonumber \\
   &&\spaEq   
   =  \sum_{n=1}^{k} 
      \left[ A_{n}(\varphi)+\varphi^{2} \right]   
      \left[ \tilde{z}_{t_{n}} 
         + \frac{\varphi}{A_{n}(\varphi)+\varphi^{2} }
            \; \tilde{z}_{t_{n+1}} \right]^{2}   
      \nonumber \\
   &&\spaEq\spaEq 
      + A_{k+1}(\varphi)\tilde{z}_{k+1}, 
%   \nonumber \\
\label{WorkCalcuSuppl1}\end{eqnarray}
for a constant $\varphi$ and $k=1,2,\cdots$, 
where $A_{n}(\varphi)$ is defined by  
\begin{eqnarray}
   A_{n}(\varphi) \equiv 
%   \frac{1}{1+\varphi^{2}+\varphi^{4}
%      +\cdots+\varphi^{2(n-1)}} 
%   =
   \frac{1}{\sum_{k=0}^{n-1}\varphi^{2k}}
   =\frac{1-\varphi^{2}}{1-\varphi^{2n}}. 
\label{FuncAn1}\end{eqnarray}
   We can prove Eq. (\ref{WorkCalcuSuppl1}) for any integer $k$ 
by mathematical induction, using the fact that  
the function $A_{n}(\varphi)$ given by Eq. 
(\ref{FuncAn1}) satisfies the recurrence formula 
\begin{eqnarray}
   A_{n+1}(\varphi) 
%   = \frac{1}{1+\varphi^{2}A_{n}(\varphi)^{-1}} 
   = \frac{A_{n}(\varphi)}{A_{n}(\varphi)+\varphi^{2}}.
\label{FuncAn2}\end{eqnarray}
   Using Eq. (\ref{WorkCalcuSuppl1}) and 
$\tilde{z}_{t_{N}}=\tilde{z}_{t}=0$ 
from Eq. (\ref{VariaZBound1}), we obtain 
\begin{eqnarray}
 &&
   \sum_{n=0}^{N-1}\left(   
      \varphi \tilde{z}_{t_{n}} 
      + \tilde{z}_{t_{n+1}} \right)^{2} 
      \nonumber \\
   &&\spaEq 
   = \sum_{n=1}^{N-1} 
      \left[ A_{n}(\varphi)+\varphi^{2} \right]   
      \left[ \tilde{z}_{t_{n}} 
         + \frac{\varphi}{A_{n}(\varphi)+\varphi^{2} }
            \; \tilde{z}_{t_{n+1}} \right]^{2} 
     \nonumber \\ 
\label{ProgTheore2}\end{eqnarray}
for a any constant $\varphi$ and $k=1,2,\cdots$. 
%-
   Using the functional integral (\ref{FunctInteg1}), 
the Lagrangian function (\ref{Lagra1}) for $v=0$, 
Eq. (\ref{ProgTheore2}) for 
$\varphi=\varphi_{N} \equiv (\Delta t_{N}/\tau)-1 $, we obtain  
\begin{widetext}
\begin{eqnarray}
   && \int_{\tilde{z}_{0}}^{\tilde{z}_{t}}
      \mathcal{D}\tilde{z}_{s} \; \exp\left[ \int_{t_{0}}^{t}ds\; 
      L^{(0)}\!\left(\dot{\tilde{z}}_{s},\tilde{z}_{s}\right)\right] 
      \nonumber \\
   &&\spaEq = 
      \lim_{N\rightarrow+\infty} 
      \left(\frac{1}{4\pi D\Delta t_{N}}\right)^{N/2}
      \int d\tilde{z}_{t_{N-1}} \int d\tilde{z}_{t_{N-2}} 
      \cdots \int d\tilde{z}_{t_{1}} 
%       \nonumber \\
%   &&\spaEq\spaEq\spaEq \times 
      \exp\left[ \sum_{n=0}^{N-1}\Delta t_{N} 
      L^{(0)}\!\left(
      \frac{\tilde{z}_{t_{n+1}}-\tilde{z}_{t_{n}}}{\Delta t_{N}},
      \tilde{z}_{t_{n}}
      \right)\right] 
      \nonumber \\
   &&\spaEq = 
      \lim_{N\rightarrow+\infty} 
      \left(\frac{1}{4\pi D\Delta t_{N}}\right)^{N/2}
      \int d\tilde{z}_{t_{N-1}} \int d\tilde{z}_{t_{N-2}} 
      \cdots \int d\tilde{z}_{t_{1}}
       %\nonumber \\
%   &&\spaEq\spaEq\spaEq \times  
      \exp\left[ -\frac{1}{4 D \Delta t_{N}} 
      \sum_{n=0}^{N-1} \left( \varphi_{N} 
      \tilde{z}_{t_{n}} + \tilde{z}_{t_{n+1}} \right)^{2}\right]
      \nonumber \\
   &&\spaEq = 
      \lim_{N\rightarrow+\infty} 
      \left(\frac{1}{4\pi D\Delta t_{N}}\right)^{N/2}
      \int d\tilde{z}_{t_{N-1}} \int d\tilde{z}_{t_{N-2}} 
      \cdots \int d\tilde{z}_{t_{1}} 
       \nonumber \\
   &&\spaEq\spaEq\spaEq \times 
      \exp\left\{ -\frac{1}{4 D \Delta t_{N}} 
      \sum_{n=1}^{N-1}
      \left[ A_{n}(\varphi_{N})+\varphi_{N}^{2} \right]   
      \left[ \tilde{z}_{t_{n}} 
         + \frac{\varphi_{N}}
            {A_{n}(\varphi_{N})+\varphi_{N}^{2} }
            \; \tilde{z}_{t_{n+1}} \right]^{2}\right\}
      \nonumber \\
   &&\spaEq = \lim_{N\rightarrow+\infty}
      \frac{1}{\sqrt{4\pi D\Delta t_{N}}} \prod_{n=1}^{N-1}
      \frac{1}{\sqrt{A_{n}(\varphi_{N})+\varphi_{N}^{2} }}
      \nonumber \\
   &&\spaEq 
   = \lim_{N\rightarrow+\infty}
      \frac{1}{\sqrt{4\pi D\Delta t_{N}}} \prod_{n=1}^{N-1}
       \sqrt{\frac{A_{n+1}(\varphi_{N})}{A_{n}(\varphi_{N})}}
      \nonumber \\
   &&\spaEq 
   = \lim_{N\rightarrow+\infty}
      \sqrt{\frac{A_{N}(\varphi_{N})}{4\pi D\Delta t_{N}}} 
      \nonumber \\
   &&\spaEq 
      = \lim_{N\rightarrow+\infty}
      \left\{2\pi D \tau\left( 1-\frac{t-t_{0}}{2\tau N}\right)^{-1}
      \left[1- \left(1-\frac{t-t_{0}}{\tau N}\right)^{2N} \right]
      \right\}^{-1/2} 
      \nonumber \\
   &&\spaEq 
      = \frac{1}{\sqrt{2\pi D \tau \left(1- b_{t}^{2} \right)}} 
%      \!\!\!\!\!\!.\;\;\;\;\;\;
\label{WorkCalcuSuppl2}\end{eqnarray}
\end{widetext}
where we used Eqs. (\ref{FuncAn1}), (\ref{FuncAn2}) 
and $\exp (X) = \lim_{N\rightarrow +\infty} (1+X/N)^{N}$ 
for any $X$.
   From Eq. (\ref{WorkCalcuSuppl2}) and   
$\mathcal{T}_{t} = (\tau/2)(1 - b_{t}^{2})$ 
we derive Eq. (\ref{PathInteg1}).

%%%%%%%%%%%%%%%%%%%%%%%%%%%%%%%%%%%%%%%%%%%%%%%%%%%%%%%%%%%%%%%%%%%%%%
%\section{Distribution Function of Energy Loss by Friction}
%\label{CalculationDistributionFriction}
%
%   In this Appendix, we give calculation details to derive  
%Eq. (\ref{DistrFrict3}). 
%and (\ref{FluctTheorFrict3}). 

%%%%%%%%%%%%%%%%%%%%%%%%%%%%%%%%%%%%%%%%%%%%%%%%%%%%%%%%%%%%%%%%%%%%%%
%\section{Calculation for the Heat Distribution}
%\label{CalculationDistributionHeat}
%
%   In this Appendix, we give calculation details of 
%Eq. (\ref{EFunctHeat3}).

%%%%%%%%%%%%%%%%%%%%%%%%%%%%%%%%%%%%%%%%%%%%%%%%%%%%%%%%%%%%%%%%%%%%%%
\section{Nonequilibrium Detailed Balance Including Inertia}
\label{NonequilibriumDetailedBalanceInertia}

   In this Appendix, we give a derivation of Eq. (\ref{DetaiBalan7}).

   Using Eq. (\ref{Lagra3}) we have 
\begin{widetext}
\begin{eqnarray}
    L^{(v)}\!\left(\ddot{y}_{s},\dot{y}_{s},y_{s}\right) 
    &=& -\frac{1}{4D}
       \left[
          -\dot{y}_{s} +\frac{1}{\tau} y_{s} 
          \pm v +\frac{m}{\alpha}\ddot{y}_{s} 
          +2\dot{y}_{s} +(1\mp 1)v
       \right]^{2} 
       \nonumber \\
%    &=& -\frac{1}{4D}
%       \left\{
%          \left(-\dot{y}_{s} +\frac{1}{\tau} y_{s}
%             \pm v +\frac{m}{\alpha}\ddot{y}_{s} \right)^{2}
%       \right.\nonumber \\
%    &&\spaEq
%          +2\left(-\dot{y}_{s} 
%          +\frac{1}{\tau} y_{s}
%          \pm v +\frac{m}{\alpha}\ddot{y}_{s} \right)
%          \left[2\dot{y}_{s} +(1\mp 1)v\right]
%       + \left[2\dot{y}_{s} 
%       + (1\mp 1)v\right]^{2}
%       \Biggr\}
%       \nonumber \\
    &=& -\frac{1}{4D}
          \left(-\dot{y}_{s} +\frac{1}{\tau} y_{s}
             \pm v +\frac{m}{\alpha}\ddot{y}_{s} \right)^{2}
          -\frac{1}{D}\left(\frac{1}{\tau} y_{s}
          \pm v +\frac{m}{\alpha}\ddot{y}_{s} \right)
          \dot{y}_{s}
       \nonumber \\
    &&\spaEq 
          -\frac{1}{D}\left(\dot{y}_{s} 
          +\frac{1}{\tau} y_{s}
          +\frac{m}{\alpha}\ddot{y}_{s} 
%          +\frac{1\pm 1}{2}v
          \right)
          \frac{1\mp 1}{2}v
       \nonumber \\
%    &=& L^{(\pm v)}\left(\ddot{y}_{s},-\dot{y}_{s},y_{s}\right)
%          -\beta\left(
%             m\ddot{y}_{s}
%             + \kappa y_{s}
%          \pm \alpha v  \right)
%          \dot{y}_{s}
%          -\beta\left(
%          m\ddot{y}_{s}
%          +\alpha\dot{y}_{s} 
%          +\kappa y_{s}
%          \right)
%          \frac{1\mp 1}{2}v
%       \nonumber \\
    &=& L^{(\pm v)}\left(\ddot{y}_{s},-\dot{y}_{s},y_{s}\right)
        -\beta\left[\left( m\ddot{y}_{s}+ \kappa y_{s} \right)
        \left(\dot{y}_{s}+\frac{1\mp 1}{2}v\right)
        + \alpha  \dot{y}_{s}\frac{1\pm 1}{2} v \right]
        \nonumber \\
    &=& L^{(\pm v)}\left(\ddot{y}_{s},-\dot{y}_{s},y_{s}\right)
          -\beta\left[
             m\ddot{y}_{s} (\dot{y}_{s} + \vartheta v)
             + \kappa y_{s} \dot{y}_{s} \right]
          +\beta m\ddot{y}_{s} \vartheta v
       \nonumber \\
    &&\spaEq 
      -\beta\left(
             m\ddot{y}_{s}
             + \kappa y_{s} \right)\frac{1\mp 1}{2}v
          - \beta \alpha  \dot{y}_{s}\frac{1\pm 1}{2} v 
       \nonumber \\
    &=& L^{(\pm v)}\left(\ddot{y}_{s},-\dot{y}_{s},y_{s}\right)
          -\beta\frac{d}{ds} \left[
             \frac{1}{2} m (\dot{y}_{s} + \vartheta v)^{2}
             + \frac{1}{2}\kappa y_{s}^{2} \right]
       \nonumber \\
    &&\spaEq   -\beta\left[
        \kappa y_{s} \frac{1\mp 1}{2} 
        + \alpha  \dot{y}_{s}\frac{1\pm 1}{2} 
        + m\ddot{y}_{s} \left(\frac{1\mp 1}{2}-\vartheta\right) 
        \right] 
       \nonumber \\
    &=& L^{(\pm v)}\left(\ddot{y}_{s},-\dot{y}_{s},y_{s}\right)
       -\beta\frac{d \mathcal{H}\!\left(
          \dot{y}_{s}+\vartheta v,y_{s}\right)}{ds} 
       +\beta\Lambda_{\pm}(\ddot{y}_{s},\dot{y}_{s},
       y_{s};\vartheta) v
\label{DetaiBalan7sup}\end{eqnarray}
\end{widetext}
where we used 
%$\tau\equiv \alpha/\kappa$ and $D\equiv 1/(\alpha\beta)$,  
Eqs. (\ref{HamilFunct1}) and (\ref{GenerForce1}), 
and $\vartheta$ is a parameter. 
   Equation (\ref{DetaiBalan7sup}) leads to 
\begin{eqnarray}
   &&
   e^{-\beta \int_{t_{0}}^{t}ds\; 
   \Lambda_{\pm}(\ddot{y}_{s},\dot{y}_{s},y_{s};\vartheta) v}
   e^{\int_{t_{0}}^{t} ds\; 
   L^{(v)}\!\left(\ddot{y}_{s},\dot{y}_{s},y_{s}\right)} 
   e^{-\beta \mathcal{H}\!\left(
   \dot{y}_{0}+\vartheta v,y_{0}\right)}
   \nonumber \\
   &&\spaEq = 
  e^{-\beta \mathcal{H}\!\left(
  \dot{y}_{t}+\vartheta v,y_{t}\right)} 
   e^{\int_{t_{0}}^{t} ds\; 
   L^{(\pm v)}\left(\ddot{y}_{s},-\dot{y}_{s},y_{s}\right)} .
\label{ModifDetaiBalanZero}\end{eqnarray}
Equation (\ref{DetaiBalan7}) is derived from 
Eq. (\ref{ModifDetaiBalanZero}) using Eq. 
(\ref{EquilDistr2}).

%%%%%%%%%%%%%%%%%%%%%%%%%%%%%%%%%%%%%%%%%%%%%%%%%%%%%%%%%%%%%%%%%%%%%%

%%%%%%%%%%%%%%%%%%%%%%%%%%%%%%%%%%%%%%%%%%%%%%%%%%%%%%%%%%%%%%%%%%%%%%
%%%%%%%%%%%%%%%%%%%%%%%%%%%%%%%%%%%%%%%%%%%%%%%%%%%%%%%%%%%%%%%%%%%%%%


\begin{thebibliography}{99} 

%--- Fluctuation-dissipation theorem

\bibitem{E05} A. Einstein, 
   Ann. Physik \textbf{17}, 549 (1905).
\bibitem{J28} J. B. Johnson, 
   Phys. Rev. \textbf{32}, 97 (1928).
\bibitem{N28} H. Nyquist, 
   Phys. Rev. \textbf{32}, 110 (1928).
\bibitem{O31a} L. Onsager, 
   Phys. Rev. \textbf{37}, 405 (1931). 
\bibitem{O31b} L. Onsager, 
   Phys. Rev. \textbf{38}, 2265 (1931).
\bibitem{C45} H. B. G. Casimir, 
   Rev. Mod. Phys. \textbf{17}, 343 (1945). 
\bibitem{G51} M. S. Green, 
   J. Chem. Phys. \textbf{19}, 1036 (1951).
\bibitem{CW51} H. B. Callen and T. A. Welton
   Phys. Rev. \textbf{83}, 34 (1951).
%\bibitem{CBJ52} H. B. Callen, M. L. Barasch and J. L. Jackson, 
%   Phys. Rev. \textbf{88}, 1382 (1952). 
\bibitem{K57} R. Kubo, 
   J. Phys. Soc. Jap. \textbf{12}, 570 (1957). 

%--- Onsager-Machlup theory
   
\bibitem{H52} N. Hashitsume, 
   Prog. Theor. Phys. \textbf{8}, 461 (1952).
\bibitem{OM53} L. Onsager and S. Machlup, 
   Phys. Rev. \textbf{91}, 1505 (1953).
\bibitem{MO53} S. Machlup and L. Onsager, 
   Phys. Rev. \textbf{91}, 1512 (1953).  
   
\bibitem{H76} H. Hasegawa, 
   Prog. Theor. Phys. \textbf{56}, 44 (1976).
\bibitem{H77} H. Hasegawa, 
   Prog. Theor. Phys. \textbf{58}, 128 (1977).
\bibitem{Y71} K. Yasue, 
   J. Math. Phys. \textbf{19}, 1671 (1978).
\bibitem{HR81} K. L. C. Hunt and J. Ross, 
   J. Chem. Phys. \textbf{75}, 976 (1981).  
\bibitem{R89} H. Risken, 
   {\it The Fokker-Planck equation : 
   methods of solution and applications}  
   (Springer-Verlag Berlin, 1989). 
   
\bibitem{BSG01} L. Bertini, A. De Sole, D. Gabrielli, 
   G. Jona-Lasinio, and C. Landim, 
   Phys. Rev. Lett. \textbf{87}, 040601 (2001).
\bibitem{BSG02} L. Bertini, A. De Sole, D. Gabrielli, 
   G. Jona-Lasinio, and C. Landim, 
   J. Stat. Phys. \textbf{107}, 635 (2002).
\bibitem{G02} G. Gallavotti, ESI preprint 1144.
   
%--- Fluctuation Theorems (1)

\bibitem{ECM93} D. J. Evans, E. G. D. Cohen, and G. P. Morriss, 
   Phys. Rev. Lett. \textbf{71}, 2401 (1993); 
   \textbf{71}, 3616 (1993) [errata]. 
\bibitem{ES94} D. J. Evans and D. J. Searles, 
   Phys. Rev. E \textbf{50}, 1645 (1994).
\bibitem{GC95} G. Gallavotti and E. G. D. Cohen, 
   Phys. Rev. Lett. \textbf{74}, 2694 (1995). 
\bibitem{K98} J. Kurchan, 
   J. Phys. A: Math. Gen. \textbf{31}, 3719 (1998).
\bibitem{LS99} J. L. Lebowitz and H. Spohn, 
   J. Stat. Phys. \textbf{95}, 333 (1999). 
\bibitem{C99} G. E. Crooks, 
   Phys. Rev. E \textbf{60}, 2721 (1999).  
\bibitem{C00} G. E. Crooks, 
   Phys. Rev. E \textbf{61}, 2361 (2000). 

%---  Fluctuation Theorems (2, experiments)

\bibitem{CL98} S. Ciliberto and C. Laroche, 
   J. Phys. IV France \textbf{8}, 215 (1998). 
\bibitem{WSM02} G. M. Wang, E. M. Sevick, E. Mittag, 
   D. J. Searles, and D. J. Evans,  
   Phys. Rev. Lett. \textbf{89}, 050601 (2002).  
%\bibitem{CRW94} D. M. Carberry, J. C. Reid, G. M. Wang, 
%   E. M. Sevick, Debra J. Searles, and Denis J. Evans, 
%   Phys. Rev. Lett. \textbf{92}, 140601 (2004). 
\bibitem{CGH04} S. Ciliberto, N. Garnier, S. Hernandez, 
   C. Lacpatia, J.-F. Pinton, G. R. Chavarria, 
   Physica A \textbf{340}, 240 (2004).
\bibitem{FM04} K. Feitosa and N. Menon, 
   Phys. Rev. Lett. \textbf{92}, 164301 (2004). 
%\bibitem{CRJ05} D. Collin, F. Ritort, C. Jarzynski, S. B. Smith, 
%   I. Tinoco Jr and C. Bustamante, 
%   Nature \textbf{437}, 231 (2005). 
\bibitem{GC05} N. Garnier and S. Ciliberto,  
   Phys. Rev. E \textbf{71}, 060101R (2005).
\bibitem{SST05} S. Schuler, T. Speck, C. Tietz, J. Wrachtrup, 
   and U. Seifert, Phys. Rev. Lett. \textbf{94}, 180602 (2005). 
   
%---  Fluctuation Theorems (3)
   
\bibitem{G96} G. Gallavotti, 
   Phys. Rev. Lett. \textbf{77}, 4334 (1996). 
   
\bibitem{ZC03a} R. van Zon and E. G. D. Cohen, 
   Phys. Rev. Lett. \textbf{91}, 110601 (2003). 
\bibitem{ZC04} R. van Zon and E. G. D. Cohen, 
   Phys. Rev. E \textbf{69}, 056121 (2004).  
   
%--- Dragged particle model

\bibitem{S98} K. Sekimoto, 
   Prog. Theor. Phys. Suppl. \textbf{130}, 17 (1998). 
\bibitem{MJ99} O. Mazonka and C. Jarzynski, 
   e-print cond-mat/9912121.
\bibitem{TTM02} S. Tasaki, I. Terasaki and T. Monnai,
   e-print cond-mat/0208154. 
\bibitem{ZC03b} R. van Zon and E. G. D. Cohen, 
   Phys. Rev. E \textbf{67}, 046102 (2003).  
%\bibitem{WSM02} G. M. Wang, E. M. Sevick, E. Mittag,  
%   D. J. Searles, and D. J. Evans,  
%   Phys. Rev. Lett. \textbf{89}, 050601 (2002).
\bibitem{ZCC04} R. van Zon, S. Ciliberto and E. G. D. Cohen, 
   Phys. Rev. Lett. \textbf{92}, 130601 (2004). 
%\bibitem{GC05} N. Garnier and S. Ciliberto, 
%   Phys. Rev. E\textbf{71}, 060101R (2005). 


%--- Note 

\bibitem{noteIIa} We note that also the equilibrium fluctuating 
equations of Onsager and Machlup have a Langevin form 
\cite{OM53,MO53}. 

%--- Comoving frame

\bibitem{TM04} T. Taniguchi and G. P. Morriss,
Phys. Rev. E \textbf{70}, 056124 (2004).

%--- note 

\bibitem{note2a}  
   Mathematically, the $v$-dependence in the Langevin equation 
(\ref{LangeEq2}) can formally be removed by 
changing the variable $y_{t}$ by $y_{t}+v\tau$. 

%--- Stochastic processes
 
\bibitem{K92} N. G. van Kampen, 
   \textit{Stochastic processes in physics and chemistry}
   (Elsevier, Amsterdam, 1992). 

%--- note 

\bibitem{noteIIA} 
   A concrete calculation process of the functional integration 
to derive Eq. (\ref{TransProba2}) from Eq. (\ref{TransProba1}) 
is similar to the one for the work distribution function 
which will be discussed in Sec. 
\ref{FunctionalCalculationWorkDistribution}. 
   More concretely, the transition probability 
$\transS{F}{y_{t}}{t}{y_{0}}{t_{0}}$ is given by 
%
%\begin{eqnarray}
$   \transS{F}{y_{t}}{t}{y_{0}}{t_{0}} 
   = \mathcal{F}(y_{t},y_{0};0) $
%\label{FF}\end{eqnarray}
%
using the function $\mathcal{F}(y_{t},y_{0}; \lambda)$ defined by 
Eq. (\ref{EFunctWorkTrnas1}), whose 
functional integral is carried out for any $\lambda$ 
in Sec. \ref{FunctionalCalculationWorkDistribution}. 


%--- Onsager-Machlup theory (II)

\bibitem{TM57} L. Tisza and I. Manning, 
   Phys. Rev. \textbf{105}, 1695 (1957). 
   
%--- Variation principle
   
\bibitem{LL69} L. D. Landau and E. M. Lifshitz, 
   \textit{Mechanics}, translated from the Russian by 
   J. B. Sykes and J. S. Bell (Pergamon Press, Oxford, 1960).

%--- note 

\bibitem{noteIIIC} More concretely, 
the solution of Eq. (\ref{EularLagra2}) 
under the conditions 
$y_{t_{0}}^{*}=y_{0}$ and $y_{t}^{*}=y_{t}$ 
is given by the case of $\lambda =0$ for 
$\ym_{s}$ in Eq. (\ref{EularLagra2Solve1}) 
which will be discussed in Sec. 
\ref{FunctionalCalculationWorkDistribution} later. 
%   The forward path (time-reversed path) 
%corresponds to the third (second) term 
%on the right-hand side of Eq. (\ref{EularLagra2Solve1}). 
 
%--- Einstein's fluctuation formula 

\bibitem{Lan59} L. D. Landau and E. M. Lifshitz, 
\textit{Statistical physics}, translated from the Russian by 
E. Peierls and R. F. Peierls 
%J. B. Sykes and M. J. Kearsley 
(Pergamon Press, London, 1958) Chapter XII.
   
%--- Path-integral Formula for the Quantum Mechanics

\bibitem{FH65} R. P. Feynman and A. R. Hibbs, 
   \textit{Quantum mechanics and path integrals} 
   (McGraw-Hill, New York, 1965).  
   
%--- Wiener Functional

%\bibitem{W23} N. Wiener, 
%   J. Math. Phys. \textbf{2}, 131 (1923). 
%\bibitem{K49} M. Kac, 
%   Transactions of the American Mathematical Society, 
%   \textbf{65}, 1 (1949). 
   
%--- note 

\bibitem{noteIVa} The asymmetry in the nonequilibrium 
detailed balance relation appears to correspond 
to the asymmetry noted by Bertini \textit{et al.} 
\cite{BSG01,BSG02} 
in the creation and decay of a fluctuation 
in a nonequilibrium steady state. 
   (In Refs. \cite{BSG01,BSG02} 
such an asymmetry is called an Onsager-Machlup symmetry, 
and we will discuss this point 
more in Eq. (\ref{OMSymme1}) 
in Sec. \ref{ConclusionsRemarks}.)
   If so, this asymmetry was applied in 
Refs. \cite{BSG01,BSG02} 
to exclusion and boundary driven zero 
range models, 
while here it applies to a stochastic model 
using the Langevin 
or the Onsager-Machlup approach. 


%---  Fluctuation Theorems (4)

\bibitem{ES02} D. J. Evans and D. J. Searles, 
   Adv. Phys. \textbf{51}, 1529 (2002).   

%--- note 
  
\bibitem{noteIIIB} 
In this paper we call the transient fluctuation theorem 
as a fluctuation theorem with the equilibrium initial condition.

%--- Identity in fluctuation theorems

\bibitem{CG99} E. G. D. Cohen and G. Gallavotti, 
   J. Stat. Phys. \textbf{96}, 1343 (1999).

%--- note 

\bibitem{noteIVc} 
   In Eq. (\ref{EFunctWorkTrnas1}) the dimensionless 
work rate is $\beta\dot{\mathcal{W}}^{(v)}(y_{s})$, 
multiplied by the Lagrange multiplier $\lambda$.
   Similarly, the third term on the right-hand side of Eq. 
(\ref{EularLagra2Modif1}) may be regarded as 
a term for the Lagrange multiplier under the restriction 
by the delta function in  Eq. 
(\ref{WorkDistr1}). 

%--- fluctuation theorems (5)

\bibitem{RCW04} J. C. Reid, D. M. Carberry, G. M. Wang, 
   E. M. Sevick, and D. J. Evans,
   Phys. Rev. E \textbf{70}, 016111 (2004). 
   
%--- Heat Fluctuation Theorem

\bibitem{BGG05} F. Bonetto, G. Gallavotti, A. Giuliani, 
   and F. Zamponi, e-print cond-mat/0507672.
\bibitem{G06} T. Gilbert, 
   Phys. Rev. E \textbf{73}, 035102R (2006).   
\bibitem{BJM06} M. Baiesi, T. Jacobs, C. Maes, and N. S. Skantzos, 
   e-print cond-mat/0602311.  

%--- Different fluctuation theorems 

\bibitem{CCJ06} V. Y. Chernyak, M. Chertkov, and C. Jarzynski, 
   e-print cond-mat/0605471.

\bibitem{ND04} O. Narayan and A. Dhar, 
    J. Phys. A: Math. Gen. \textbf{37}, 63 (2004).
\bibitem{K05} K. Kitahara, a private communication. 
\bibitem{IP06} A. Imparato and L. Peliti,  
   e-print cond-mat/0603506.


%----

\bibitem{noteHeat}  
   It should be noted that the extended heat fluctuation 
theorem may also depend on the initial condition. 
   [Note that in our simple argument for heat 
[based on Eq. (\ref{HeatDistr3}), etc.]
in the second half of Sec. \ref{FluctuationTheoremHeat} 
we assumed a canonical-like distribution 
%the equilibrium or the non-equilibrium steady state 
as the initial distribution.] 
   Actually, if we could choose the initial distribution 
$f(y_{0},t_{0})$ as a constant [although in this case 
$f(y_{0},t_{0})$ cannot be normalized 
for $y_{0}\in(-\infty,+\infty)$] then we can show that 
the heat satisfies the conventional form of fluctuation 
theorem $P_{q}(Q,t)/P_{q}(-Q,t) = \exp(Q)$ 
for any time, which is derived from Eq. (\ref{DetaiBalan6}) 
for the case of $f_{ref}(y_{0})$ to be constant, or from 
Eq. (\ref{EFunctHeat3}) leading to the relation 
$\mathcal{E}_{q}(\lambda,t) = \mathcal{E}_{q}(1-\lambda,t)$ 
in this case. 

\end{thebibliography}
\end{document}